\documentclass[preprint,12pt]{elsarticle}

\usepackage{graphicx}

\def\be{\begin{equation}} 
\def\ee{\end{equation}} 
\def\bea{\begin{eqnarray}} 
\def\eea{\end{eqnarray}} 

\usepackage{amssymb}

\usepackage[nodots,nocompress]{numcompress}

\biboptions{super}

\journal{Stud. in History and Philosophy of Modern Physics}

\begin{document}

\begin{frontmatter}



\title{Do we have a Theory of Early Universe Cosmology?}


\author{Robert Brandenberger}

\address{Physics Department, McGill University, 3600 University Street, Montreal, QC, H3A 2T8, Canada}

\begin{abstract}

The inflationary scenario has become the paradigm of early universe cosmology, and -
in conjuction with ideas from superstring theory - has led to speculations about an
``inflationary multiverse". From a point of view of phenomenology, the inflationary
universe scenario has been very successful. However, the scenario suffers from some 
conceptual problems, and thus it does not (yet) have the status of a solid theory. 
There are alternative ideas for the evolution of the
very early universe which do not involve inflation but which agree with most current
cosmological observations as well as inflation does. In this lecture I will outline
the conceptual problems of inflation and introduce two alternative pictures -
the ``matter bounce" and ``string gas cosmology", the latter being a realization
of the ``emergent universe" scenario based on some key principles of superstring
theory. I will demonstrate that these two alternative pictures lead to the same
predictions for the power spectrum of the observed large-scale structure and for
the angular power spectrum of cosmic microwave background anisotropies as the inflationary
scenario, and I will mention predictions for future observations with which the
three scenarios can be observationally teased apart.

\end{abstract}


\end{frontmatter}


\section{Introduction}
\label{section:Intro}

Inflationary cosmology was proposed \cite{Guth} (see also \cite{Brout, Starob, Sato} for
some earlier related work) in 1981 as a solution of some
conceptual problems of the previous early universe paradigm, the
Standard Big Bang model. It was soon realized \cite{Mukh} (see also \cite{Sato, Press}
for related insights) that inflationary cosmology leads to the first causal theory for
the origin of the structure in the universe which we measure today in
terms of inhomogeneities in the distribution of galaxies on large scales,
and anisotropies in cosmic microwave background (CMB) temperature maps. The
predictions of this theory have been spectacularly confirmed by
observations, in particular by the WMAP satellite results for anisotropies
in the CMB \cite{WMAP}. 

Inflationary cosmology has now become the paradigm of early universe
cosmology. In most papers on early universe cosmology the impression is
given that inflation has been firmly established and observationally proven.
However, as I will argue here, this inflationary cosmology is far from proven.
Although phenomenologically successful, current realizations of inflation
suffer from serious conceptual problems, and hence inflationary cosmology
does not (yet) have the status of an established theory. Furthermore,
as realized ten years before the development of inflationary cosmology,
any theory which provides an approximately scale-invariant spectrum
of almost adiabatic primordial fluctuations in the early universe on
scales which are larger than the Hubble radius \footnote{The meaning of
the Hubble radius will be discussed later in the text.} will lead to
inhomogeneities and anisotropies in the late time universe which are
in agreement with current data \cite{Rashid, Peebles}. Although inflation
was the first theory based on causal physics to generate such a spectrum
of fluctuations, in the mean time other scenarios have been proposed
which yield the same type of spectrum. I will present two examples
\footnote{However, it must be emphasized that obtaining a scale-invariant
spectrum of curvature fluctuations is by no means a general outcome
of possible early universe scenarios. Pure initial vacuum and pure
thermal fluctuations in standard cosmology, for example, do not
lead to a scale-invariant spectrum.}.

In recent years, ideas from inflationary cosmology have been combined
with ideas from stochastic dynamics and from string theory to generate
the picture of the {\it inflationary multiverse} and the {\it cosmological landscape}.
In light of the problems of inflation and other concerns which will be
detailed below, I consider these ideas to be premature and warn readers
(in particular those from the philosophy of physics community) not to
consider them as established.

The outline of this article is as follows: I first discuss inflationary cosmology
and its problems. Then, I introduce two alternative scenarios for early
universe cosmology which do not involve inflation. In Section 5 (the only
technical part of this article) I focus on the development of fluctuations
in early universe cosmology, and I show
how both inflationary cosmology and the two alternative cosmologies
which I discuss yield fluctuations in agreement with current observations.
Section 6 contains some concluding remarks.

\section{Review of Inflationary Cosmology}
\label{section:infl}

The inflationary universe scenario is based on the assumption that there was
a period in the very early universe during which space expanded almost
exponentially \footnote{In fact, all that is required is that the cosmological
scale factor $a(t)$ is an accelerating function of time. Most models of inflation
yield, however, almost exponential expansion. Hence, we will formulate
the arguments in this article in the context of exponential expansion.}
(see Figure 1 for a space-time sketch of inflationary cosmology).
The exponential expansion of space leads to a solution of
several problems of Standard Big Bang cosmology. First of all, it 
leads to a {\it horizon} (defined as the
forward light cone of an initial event in space-time) which is exponentially
larger than it would be without the period of inflation. Provided that
the period of inflation is sufficiently long, the horizon at the time $t_{rec}$ 
of recombination, the time when the CMB last scattered, is larger than
distance at time $t_{rec}$ which the CMB probes. The exponential
expansion of space also homogenizes the universe since the inhomogeneities
are redshifted. Hence, inflation provides an explanation for the observed
near isotropy of the CMB, i.e. it solves the {\it horizon problem} of Standard
Cosmology. Provided that the period of inflation is longer than around 
$50 H^{-1}$ (where $H$ is the expansion rate of space during the inflationary
phase), inflation can also explain the observed spatial flatness of
the universe, i.e. it solves the {\it flatness problem}. Furthermore,
the exponential expansion of space explains how a Planck-scale
initial universe at the Planck time can grow to encompass the entire
observed universe. Related to this point, the entropy production which
is a crucial aspect of inflation and occurs at the end of the inflationary
phase can explain how the presently observed large entropy of the
universe emerges. 
 
Most importantly, however, inflationary cosmology provided the first
causal theory for the origin of the structure in the universe which we
observe in terms of inhomogeneities in the distribution of galaxies
and small amplitude anisotropies in the CMB \cite{Mukh}. At the
time when this was realized, no CMB anisotropies had been observed.
It would take a decade before the initial detection \cite{COBE}, five more
years before the characteristic oscillations in the angular power spectrum
of these anisotropies were seen \cite{Boomerang}, and five more years before the
current precision maps became available \cite{WMAP}. The fact that
inflationary cosmology made predictions in agreement with the
high precision observations (after adjusting a small number of
cosmological parameters) is a major success of the scenario.
However, it is important to keep in mind that, as discussed ten
years before the development of inflationary cosmology, any
early universe model which generates a nearly {\it scale-invariant}
spectrum of primordial almost adiabatic fluctuations will be in
agreement with current observations \footnote{Scale-invariance
of the fluctuation spectrum will be defined in Section 5.} Whereas inflationary
cosmology yielded the first causal mechanism for producing
this type of fluctuations, there are now several alternative scenarios,
two of them which will be discussed in this lecture.

\begin{figure} 
\includegraphics[height=15cm]{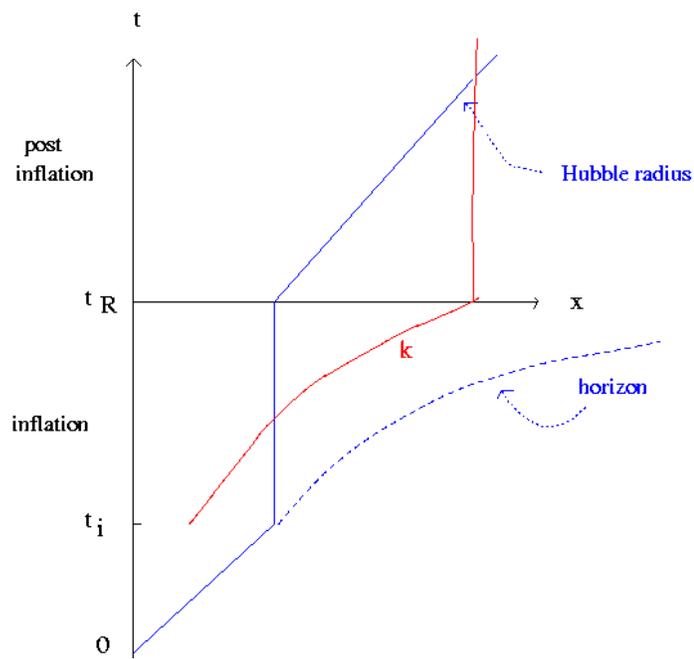}
\caption{Space-time sketch of inflationary cosmology.
The vertical axis is time, the horizontal axis corresponds
to physical distance. The solid line labelled $k$ is the
physical length of a fixed comoving fluctuation scale. The
role of the Hubble radius and the horizon are discussed
in the text.}
\label{infl1}
\end{figure}

If we use Einstein gravity to describe space and time, then exponential
expansion of space requires matter to be dominated by some substance
with an equation of state $p = - \rho$, where $p$ and $\rho$ are
pressure and energy densities, respectively. Matter which we observe
in nature does not have such highly negative pressure. Even if we
describe matter in terms of quantum fields (as we should since quantum
field theory is the best description we have which describes matter at
very high energies and thus should be used in early universe cosmology),
obtaining the above equation of state requires assuming the existence
of a {\it scalar} matter field $\varphi$ with a non-vanishing potential
energy density $V(\varphi)$. This potential energy leads to a contribution
to energy density and pressure which is in agreement with the above
equation of state.

The inflationary scenario now is as follows \cite{Linde}: at the initial time, the
scalar field is displaced from the minimum of its potential. Since the
potential is tuned to be very flat, the scalar field motion is very slow.
Thus, the scalar field potential energy density remains almost constant,
whereas all other forms of matter redshift \footnote{The inflationary slow-roll 
trajectory is a local attractor in initial condition space \cite{attractor}.} .
Thus, at some time $t_i$,
the scalar field potential energy starts to dominate and inflation begins.
Once the scalar field has decreased to a critical value which in many
models is close to the Planck scale, the scalar field kinetic energy begins
to dominate over the potential energy and inflation ends. The time when
this occurs is denoted as $t_R$.

No fundamental particles of nature which correspond to scalar fields
have yet been observed in nature. The Standard Model of particle physics
does contain one such field, the {\it Higgs field}, the field responsible
for giving the particles we observe masses. Tantalizing evidence for
the existence of the Higgs field is emerging from the recent experiments
at CERN, but the evidence is not yet strong enough to claim a discovery.
Unfortunately, the potential of the Higgs field has the wrong shape and
cannot yield cosmological inflation, unless a very unusual coupling of
this field to gravity is assumed \cite{Shaposh}. The scalar field potential
must be tuned to be very flat in order to allow for a long period of time
during which the potential energy of the field dominates over other forms.

Theories beyond the Standard Model of particle physics often contain
scalar fields. In particular, supersymmetric theories have scalar field
partners to all fundamental fermionic fields of spin 1/2. However,
to obtain the required flat potentials for these scalar fields requires
fine-tuning. Thus, although cosmologists now have many toy models
for inflation, there is no convincing realization of the scenario. 
There are also attempts to obtain inflation from modified theories of
gravity rather than by introducing new forms of matter. In fact,
the first model of exponential expansion of space \cite{Starob} was
based on a modified gravity model. However, also in this context
inflation does not seem to emerge in a natural way.

In spite of the lack of a firmly established theory of inflation,
inflationary model building has become an industry (see e.g.
\cite{inflrevs} for reviews). Stochastic effects due to quantum
fluctuations have been added \cite{Starob2} to the classical
scalar field evolution, leading to the conclusion that in
some regions of field space there is a finite probability that
the scalar field will move up the potential rather than down
as it does with classical dynamics alone. This gives rise to
a scenario \cite{eternal} in which inflation is eternal into 
the future: most of the physical volume of space remains in a state 
of inflation, and only pockets of space can exit inflationary expansion and
can make the transition to the Standard Big Bang phase. These 
are pockets of space where the scalar field value has dropped
below the critical value for stochastic effects to be
important. The fancy name for this scenario is the {\it eternal
self-reproducing inflationary universe}.

A further twist to this scenario resulted from merging the ideas
of stochastic inflation with the realization that the ground
state of string theory is not unique. Physicists now talk about
an enormous number of {\it string vacua}, ground states of
string theory \cite{number}. Each of these ground states may have a different
value of the cosmological constant $\Lambda$, and one may appeal
to anthropic arguments to claim that the universe will select
the ground state with exactly the observed value of $\Lambda$.
The fancy name for the resulting picture is the {\it string
inflationary multiverse} \cite{multiverse} or the {\it string
landscape}.

There are a couple of points to make with respect to the above
recent ideas. First of all, there is nothing special to string
theory in terms of having many ground states. Imagine a
geographic landscape: it will also have many local minima
of the gravitational potential. Second, the current analysis
of string vacua is based on quantum field theory ideas inspired
by string theory, not by non-perturbative string theory. We
are still lacking a non-perturbative approach to string theory
which is consistent with a positive or vanishing cosmological
constant. There is in fact some evidence that many
ground states of a string theory-motivated effective field 
theory are not realizable in full string theory \cite{swamp}.
Thirdly, the mathematical analysis which tries to include
stochastic effects and which leads to the eternal self-reproducing 
universe is not under good mathematical controle. In
fact, back-reaction effects \cite{back} which are
not included in stochastic inflation may invalidate
the entire analysis.

\section{Challenges for Inflationary Cosmology}

We have already discussed the first problem of inflaton, 
namely the question of how to embed inflation into a
natural particle physics model. If we manage to overcome
this problem, a second problem immediately rears its head,
namely the {\bf amplitude problem}. The inflationary
scenario was introduced in order to eliminate fine tuning
conditions for the initial data set of cosmology, and it 
provides a mechanism for the origin of structure in the universe.
However, in a wide class of inflationary
models, obtaining the correct amplitude of the cosmological
perturbations requires the introduction
of a large hierarchy in scales \cite{Adams}, and thus the
fine-tuning problem returns in a different guise. 

 \begin{figure}
\includegraphics[height=15cm]{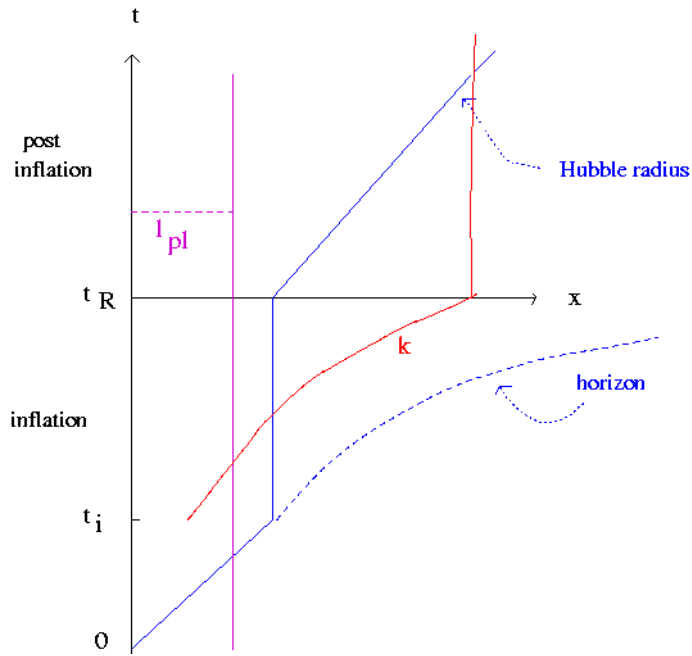}
\caption{Space-time diagram (sketch) 
of inflationary cosmology where we have added an extra
length scale, namely the Planck length $l_{pl}$ (majenta vertical line).
The symbols have the same meaning as in Figure 2.
Note, specifically, that - as long as the period of inflation
lasts a couple of e-foldings longer than the minimal value
required for inflation to address the problems of Standard
Big Bang cosmology - all wavelengths of cosmological interest
to us today start out at the beginning of the period of inflation
with a wavelength which is smaller than the Planck length.}
\label{infl2}       
\end{figure}

A more serious problem is the {\bf trans-Planckian problem} 
for fluctuations \cite{RHBrev3}. Let us return to 
the space-time diagram of inflation (see Figure \ref{infl2}).
The success of inflation at providing a causal structure
formation mechanism is based on the fact that scales
which are currently observed originate with a physical 
wavelength smaller than the Hubble radius at the beginning
of the period of inflation. This typically requires inflation
to last about $50 H^{-1}$, where $H$ is the expansion rate
during the inflationary phase. However, if the period of
inflation was only a bit longer, namely longer than about
$70 H^{-1}$, then the wavelengths of all currently observable
scales were in fact smaller than the Planck length at the 
initial time of inflation. The problem is that we do not
know the physics operative on these scales. It is clear
that both Einstein's theory of General Relativity and the theory
of scalar field matter (the two ingredients on which the
current theory of cosmological perturbations is based)
will badly break down on these scales. Thus, the key
successful prediction of inflation (the theory of the origin of
fluctuations) is based on suspect calculations since 
new physics {\it must} enter into a correct computation of the 
spectrum of cosmological perturbations. The key question is as 
to whether the predictions obtained using
the current theory are sensitive to the specifics of the unknown
theory which takes over on small scales. Under certain
assumptions about how physics at very high energy
scales communicates with low energy physics it
can be shown that the results are insensitive
(see e.g. \cite{Shenker} for an early work). However,
these assumptions may not be realized in nature.
In fact, simple toy models of new physics on super-Planck scales 
based on modified dispersion relations \cite{Jerome1} (see also
\cite{Niemeyer}) or pre-inflationary dynamics \cite{Zhang}
show that the resulting spectrum of
cosmological fluctuations depends in an important
way on what is assumed about physics on trans-Planckian scales.
If the inflationary phase lasts for less than $70 H^{-1}$,
then the trans-Planckian uncertainties only effect fluctuations
on smaller length scales (length scales on which the
structure of the universe today is non-linear) and thus
they are harder to test in cosmological precision experiments. 
 
A fourth problem is the {\bf singularity problem}. It was known for a long
time that Standard Big Bang cosmology cannot be the complete story of
the early universe because of
the initial singularity, a singularity which is unavoidable when basing
cosmology on Einstein's field equations in the presence of a matter
source obeying the weak energy conditions (see e.g. \cite{HE} for
a textbook discussion). The singularity theorems have been
generalized to apply to Einstein gravity coupled to scalar field
matter, i.e. to scalar field-driven inflationary cosmology \cite{Borde}.
It was shown that, in this context, a past singularity at some point
in space is unavoidable. Thus we know, from the outset, that scalar
field-driven inflation cannot be the ultimate theory of the very
early universe.

The Achilles heel of scalar field-driven inflationary cosmology may
be the {\bf cosmological constant problem}. We know from
observations that the large quantum vacuum energy of field theories
does not gravitate today. However, to obtain a period of inflation
one is using the part of the energy-momentum tensor of the scalar field
which looks like the vacuum energy. In the absence of a 
solution of the cosmological constant problem it is unclear whether
scalar field-driven inflation is robust, i.e. whether the
mechanism which renders the quantum vacuum energy gravitationally
inert today will not also prevent the vacuum energy from
gravitating during the period of slow-rolling of the inflaton 
field.

A final problem which we will mention here is the concern that the
energy scale at which inflation takes place is too high to justify
an effective field theory analysis based on Einstein gravity. In
simple toy models of inflation, the energy scale during the period
of inflation is about $10^{16} \rm{GeV}$, very close to the string scale
in many string models, and not too far from the Planck scale. Thus,
correction terms in the effective action for matter and gravity may 
already be important at the energy scale of inflation, and the 
cosmological dynamics may be rather different from what is
obtained when neglecting the correction terms.

\begin{figure}
\includegraphics[height=15cm]{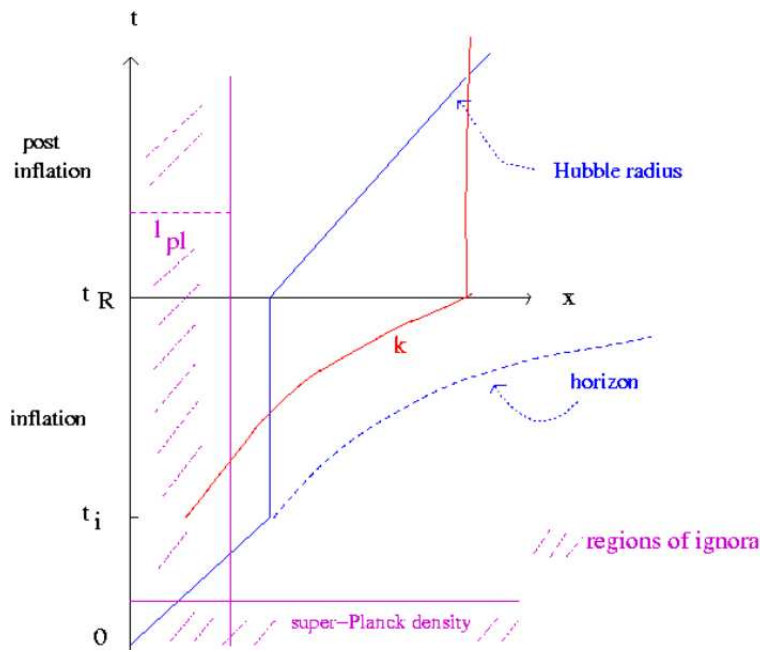}
\caption{Space-time diagram (sketch) 
of inflationary cosmology including the two zones of
ignorance - sub-Planckian wavelengths and trans-Planckian 
densities. The symbols have the same meaning as in Figure 2.
Note, specifically, that - as long as the period of inflation
lasts a couple of e-foldings longer than the minimal value
required for inflation to address the problems of Standard
Big Bang cosmology - all wavelengths of cosmological interest
to us today start out at the beginning of the period of inflation
with a wavelength which is in the zone of ignorance.}
\label{infl3}       
\end{figure}
 
In Figure \ref{infl3} we show once again the space-time sketch
of inflationary cosmology. In addition to the length scales
which appear in the previous versions of this figure,
we have now shaded the ``zones of ignorance", zones
where the Einstein gravity effective action is sure to
break down. As described above, fluctuations emerge
from the short distance zone of ignorance (except if
the period of inflation is very short), and the energy
scale of inflation might put the period of inflation too
close to the high energy density zone of ignorance to
trust the predictions based on using the Einstein
action. 

Because of the above conceptual problems, the inflationary
scenario may be the current paradigm of early uiverse cosmology,
but it is by no means a complete theory. The problems
provide a strong motivation to search for alternative early
universe scenarios, two of which are described below. To be
of interest, an alternative scenario should solve most of the
problems of Standard Big Bang cosmology which inflation
addresses, it should mitigate some of the problems of
inflation, it must be consistent with the current data on
the large-scale structure of the universe, and it should
make some predictions with which it can be distinguished
with future observations from the inflationary scenario.

\section{Two Alternative Scenarios}
\label{section:Alt}

\subsection{Matter Bounce}

The idea of a bouncing universe has a long history both in physics
and in cosmology. In such a scenario, time is eternal both in our past
and in our future. This has implications for the philosophy of cosmology:
there is no mysterious creation event, and we are no longer faced with
the problem - insoluble in the context of pure physics - of explaining
an initial singularity \footnote{However, even if the creation event
is absent, there are nevertheless conceptual/philosophical problems.
For a discussion of this issue see e.g. \cite{HZ1}.}.

There have been many attempts to construct bouncing cosmologies
in the context of physical cosmology (for references the reader is
referred to \cite{Novello}). More recently, there have been attempts
to use bouncing cosmologies to provide mechanisms alternative to
inflation for providing the scale-invariant spectrum of cosmological
observations which observations tell us must have been generated
in the early universe. The ``Pre-Big-Bang" scenario \cite{PBB}
is one attempt which was motivated from string theory dualities 
(similar dualities as the dualities which give rise to the ``String Gas
Cosmology" \cite{BV} implementation of the ``emergent Universe" scenario
which will be discussed in the second part of this section). A more
recent construction is the ``Ekpyrotic Universe" scenario \cite{Ekp}
(and its cyclic version \cite{Ekpcyclic}) which was initially also
motivated by superstring theory. Neither the Pre-Big-Bang nor
the Ekpyrotic scenario provide a scale-invariant spectrum of
adiabatic fluctuations (at least when analyzed from a purely
four space-time dimensional point of view - see \cite{Thorsten}
for a higher-dimensional study), but it is not hard to find variants
of the scenarios which lead to a scale-invariant spectrum of
entropy fluctuations (which then later convert to adiabatic
fluctuations).

I wish to discuss a simple bouncing scenario which - as realized
in \cite{Wands, Fabio2} - leads to a scale-invariant spectrum of
density perturbations. This is the ``matter bounce" scenario.
Roughly speaking, the scenario involves a contracting
phase which is the time reverse
of our current cosmological phase of expansion which is then
smoothly connected at some high density to the expanding
phase of Standard Cosmology. Obviously, new physics
is required to obtain a smooth transition from contraction to
expansion. 

{F}igure 4 is a space-time sketch of the matter bounce
scenario. The vertical axis is time, the horizontal axis
corresponds to comoving length. As is obvious, modes
which we observe today start out on sub-Hubble scales
in the early phase of contraction. Hence, it is possible
to have a causal mechanism for structure formation. As will
be discussed in the section on cosmological perturbations,
we assume that perturbations start out in their vacuum state.
This is in line with the assumption that the universe started
out large and almost empty. What it crucial for the success of the matter bounce
scenario as a theory of structure formation is that wavelengths which we 
observe today in cosmological observations left the Hubble radius 
during the matter-dominated phase of contraction. It is these
scales which acquire a scale-invariant spectrum.
 
\begin{figure}[htbp] 
\includegraphics[height=7cm]{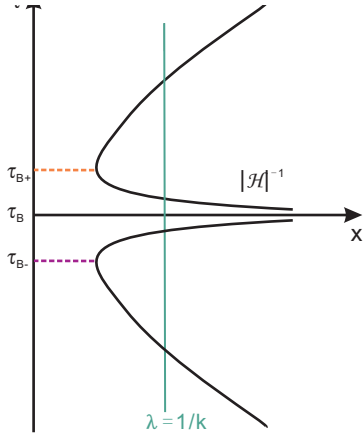}
\caption{Space-time sketch in the matter bounce scenario. The vertical axis
is conformal time $\eta$, the horizontal axis denotes a co-moving space coordinate
(a co-moving coordinate grid expands as space expands like a grid painted onto
the surface of a balloon expands as the balloon is inflated).
The vertical line indicates the wavelength of a fluctuation mode.
Also, ${\cal H}^{-1}$ denotes the co-moving Hubble radius.}
\label{bounce}
\end{figure}

How are the problems of Standard Big Bang cosmology (discussed
at the beginning of Section 2) addressed in
the matter bounce scenario? First of all, note that since
the universe begins large, the size and entropy problems 
of Standard Cosmology do not arise.
In addition, there is no horizon problem
as long as the contracting period is long
(to be specific, of similar duration as the post-bounce expanding
phase until the present time). As already mentioned, there
is a causal mechanism for generating the primordial
cosmological perturbations which evolve into the structures we observe
today. 

The flatness problem is
the one which is only partially addressed in the matter bounce
setup. The contribution of the spatial curvature decreases
in the contracting phase at the same rate as it increases in
the expanding phase. Thus, to explain the observed spatial
flatness, comparable spatial flatness at early times in the
contracting phase is required. This is an improved situation compared
to the situation in Standard Big Bang cosmology where spatial
flatness is overall an unstable fixed point and hence extreme
fine tuning of the initial conditions is required to explain
the observed degree of flatness. But the situation is not as
good as it is in a model with a long period of inflation where spatial
flatness is a local attractor in initial condition space (it is not
a global attractor, though!).

As any good alternative to the inflationary scenario should,
the matter bounce paradigm successfully addresses some of
the conceptual problems of inflation. Possibly most importantly, 
the length scale of fluctuations of interest for current observations 
on cosmological scales is many orders of magnitude larger than 
the Planck length throughout the evolution. If the energy scale at the
bounce point is comparable to the particle physics 
``Grand Unified Theory" (GUT)
scale, then typical wavelengths at the bounce point are
not too different from $1 {\rm mm}$. Hence, the fluctuations
never get close to the small wavelength zone of ignorance in
Figures 2 and 3, and thus a description of the evolution of
fluctuations using Einstein gravity should be well justified
modulo possible difficulties at the bounce point which we will
return to. Thus, there is no trans-Planckian problem
for fluctuations in the matter bounce scenario.

As already mentioned, new physics is required in order to
provide a non-singular bounce. Thus, the ``solution" of the
singularity problem is put in by hand and cannot be counted
as a success, except in realizations of the matter bounce in
the context of a string theory background in which the non-singular
evolution follows from general principles. Such a 
theory has recently been presented in \cite{KPT} (see
\cite{BKPPT} for an analysis of fluctuations in these models).
Existing matter bounce models do not address the cosmological
constant problem. However, I would like to emphasize that
the mechanism which drives the evolution in the matter bounce
scenario is robust against our ignorance of what solves the
cosmological constant problem, an improvement of the
situation compared to inflationary cosmology.

However, the matter bounce scenario presented here suffers
from a serious anisotropy problem. Since the energy density
in anisotropies scales as $a^{-6}$, where $a(t)$ is the cosmological
scale factor, and thus increases faster than that of matter (scaling
as $a^{-3}$) and radiation (scaling as $a^{-4}$), it increases
relative to that of matter and radiation. This endangers the
smooth nearly homogeneous and isotropic bounce, unless the
initial anisotropy is tuned to an extremely small value.
Note that the Ekpyrotic scenario, which also involves a phase
of contraction, is immune from this problem since the
energy density in the field which is responsible for the Ekpyrotic
contraction grows even faster as $a$ decreases.

With Einstein gravity and matter satisfying the usual energy conditions
it is not possible to obtain a non-singular bounce. Thus, new
physics is required in order to obtain a non-singular bouncing
cosmology. Such new physics can arise by modifying either the 
gravitational or the matter sector. 

To obtain a bouncing cosmology by modifying matter it is necessary 
to introduce a new form of matter which violates the 
Null Energy Condition (NEC), a condition which prohibits
negative energy densities. It is necessary to arrange this
new form of matter such that it dominates only at very
high densities. The simplest way to obtain a bounce is
to use two forms of matter, the first regular matter,
and the second a ``quintom field" \cite{quintom}
whose contributions to the energy density and pressure are
opposite to that of regular matter. The universe
starts out in a contracting phase dominated by regular
matter. The dynamics of the quintom field is then arranged 
in a way such that at high energy densities the absolute
value of the energy density in the quintom field increases
compared to that in regular matter. The total energy
density then decreases to zero, allowing for a cosmological
bounce to occur \cite{Cai1} (some important details are omitted in
this discussion, and the interested reader is referred to
\cite{RHBbouncerev} for details). This simplest way of obtaining a
cosmological bounce is plagued by an instability
of the vacuum \cite{ghost}. More sophisticated models
which avoid this instability have recently been developed,
e.g. the ghost condensate scenario \cite{Creminelli, Chunshan}
or the Galileon bounce \cite{GBounce}. However, these
models do not (yet) come from a theory of matter and
gravity which is complete at high energies. 

Possibly a more promising approach to obtaining bouncing
cosmologies is by modifying the gravitational sector of the
theory. There is excellent motivation to consider modifications
of gravity at high energy densities: General Relativity is not
a renormalizable quantum theory of gravity. In all known
approaches to quantum gravity, the Einstein action of
General Relativity is only a low energy effective action. At
high energy densities where the bounce is expected to
occur deviations from General Relativity will be important.
Two examples of modifications of General Relativity at
high densities which lead to bouncing cosmologies are
the ``nonsingular universe" construction of \cite{BMS}
and the ghost-free higher derivative action of \cite{Biswas}.
It was also realized \cite{HLbounce} that the
Ho\v{r}ava-Lifshitz proposal for a power-counting
renormalizable theory of quantum gravity \cite{Horava}
leads to a bouncing homogeneous and isotropic cosmology,
provided that the spatial curvature is non-vanishing.

Bouncing cosmologies also are predicted in more
ambitious approaches to quantizing gravity such
as string theory (see e.g. \cite{KPT}) and loop quantum
cosmology (see e.g. \cite{LQC} for recent reviews).

\subsection{Emergent Universe}

\begin{figure}
\includegraphics[height=6cm]{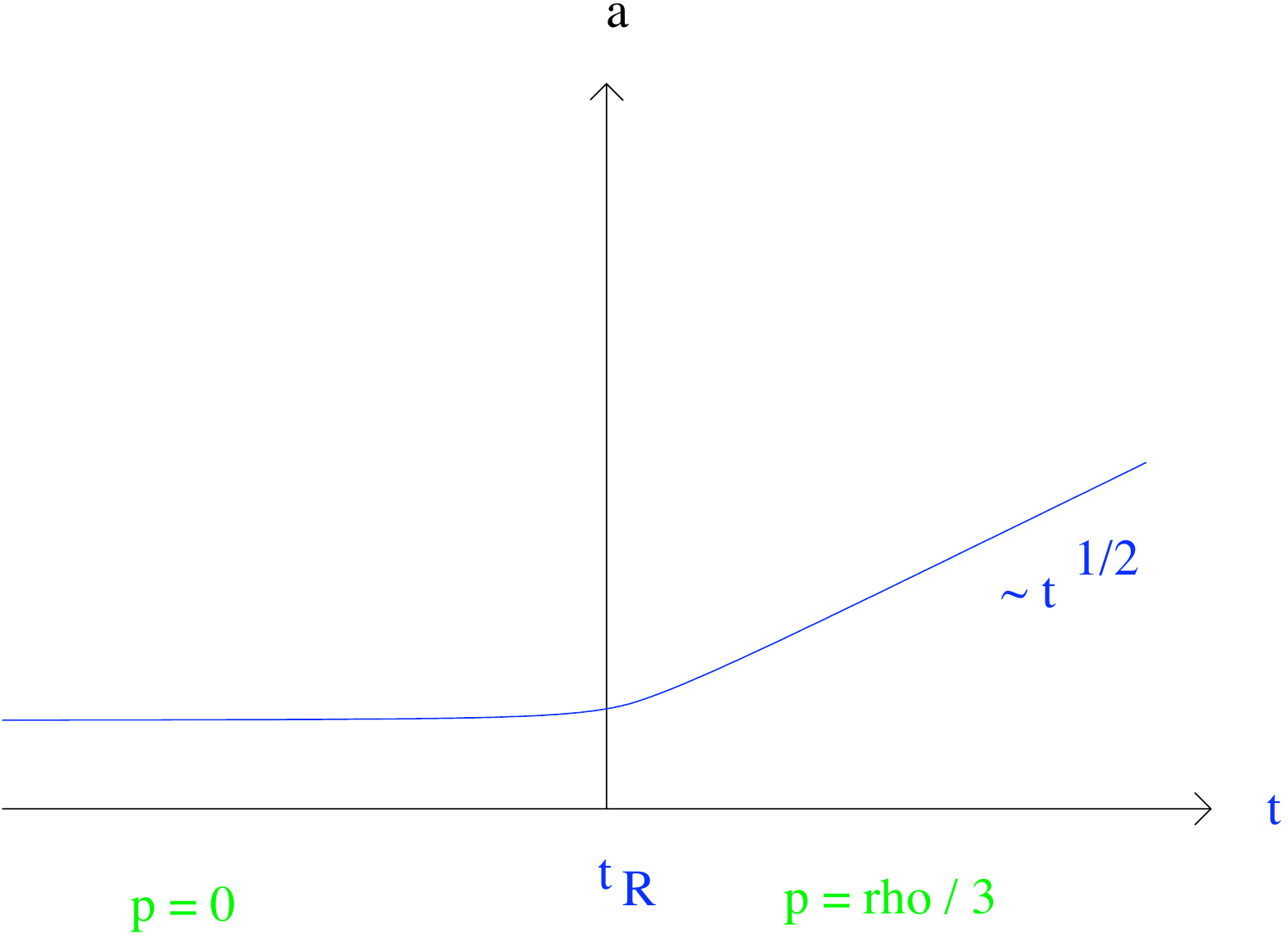}
\caption{The dynamics of emergent universe cosmology. The vertical axis
represents the scale factor of the universe, the horizontal axis is
time. }
 \label{timeevol}
\end{figure}

The ``emergent universe" scenario \cite{emergent} is another
non-singular cosmological scenario in which time runs
from $- \infty$ to $+ \infty$. The universe is assumed
to emerge in a quasi-static high density phase which at
some time (which is conventionally called $t = 0$ undergoes
a phase transition to the expanding phase of Standard
Big Bang cosmology. The time
evolution of the scale factor is sketched in Figure \ref{timeevol}.
For the purposes of establishing a theory of cosmological
structure formation, the quasi-static phase is not required
to be infinite. All that is required is that the phase is 
much longer than the length which in the expanding phase grows to 
become the current Hubble radius. The quasi-static phase
could thus be the bounce phase of a bouncing cosmology, as in
the model of \cite{Tirtho2} (which is based on the higher
derivative gravitational Lagrangian of \cite{Biswas}). 

The time evolution of the cosmological scale factor in an
emergent universe is sketched in Fig. \ref{timeevol}.
The vertical axis is the cosmological scale factor, the
horizontal axis is time. The universe is initially
static and makes a smooth transition to the radiation phase
of Standard Big Bang cosmology.

The emergent scenario is similar to inflationary cosmology in that the
universe is assumed to begin hot and small. But it is similar to a
bouncing cosmology in that time runs from $- \infty$ to $+ \infty$,
and in that the evolution is non-singular.

\begin{figure} 
\includegraphics[height=9cm]{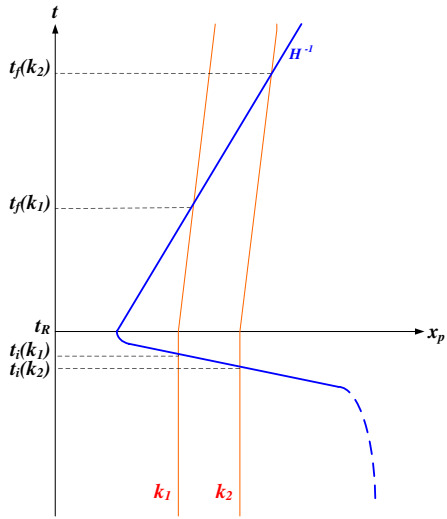}
\caption{Space-time diagram (sketch) showing the evolution of fixed 
co-moving scales in emergent cosmology. The vertical axis is time, 
the horizontal axis is physical distance.  
The solid curve represents the Hubble radius 
$H^{-1}$ which shrinks abruptly to a micro-physical scale at $t_R$ and then 
increases linearly in time for $t > t_R$. Fixed co-moving scales (the 
dotted lines labeled by $k_1$ and $k_2$) which are currently probed 
in cosmological observations have wavelengths which were smaller than 
the Hubble radius long before $t_R$. They exit the Hubble 
radius at times $t_i(k)$ just prior to $t_R$, and propagate with a 
wavelength larger than the Hubble radius until they re-enter the 
Hubble radius at times $t_f(k)$.}
\label{spacetimenew}
\end{figure}

In Figure \ref{spacetimenew} we sketch the space-time diagram
in an emergent cosmology. Since the early emergent phase is
quasi-static, the Hubble radius is infinite. For the same reason,
the physical wavelength of fluctuations remains constant in
this phase. At the end of the emergent phase, the Hubble radius
decreases to a microscopic value and makes a transition to
its evolution in Standard Cosmology.

As in inflationary cosmology and in a bouncing cosmology 
we see that fluctuations originate on sub-Hubble scales.
In emergent cosmology, it is the existence of a quasi-static phase
which leads to this result. What sources fluctuations depends on
the realization of the emergent scenario. String Gas Cosmology
is the example which I will consider later on. 
In this case, the source of perturbations 
is thermal: string thermodynamical fluctuations in
a compact space with stable winding modes, and this
in fact leads to a scale-invariant spectrum \cite{NBV}.

How does emergent cosmology address the problems of
Standard Cosmology? As in the case of a bouncing 
cosmology, the horizon is infinite and hence there is
no horizon problem. Since there is likely thermal
equilibrium in the emergent phase, a mechanism
to homogenize the universe exists. As discussed in the previous
paragraph, there is no causality obstacle against
producing cosmological fluctuations. The scenario
is non-singular, but this cannot in general be weighted
as a success unless the emergent phase can be shown
to arise from some well controlled ultraviolet physics.

Like in the case of a bouncing cosmology, there is no
trans-Planckian problem for fluctuations - their
wavelength never gets close to the Planck scale.
And like in the case of a bouncing cosmology, the
physics driving the background dynamics is
robust against our ignorance of what solves
the cosmological constant problem. These are
two advantages of the emergent scenario compared
to inflation. 

On the negative side, the origin of the large
size and entropy of our universe remains a mystery
in emergent cosmology. Also, the physics yielding
the emergent phase is not well understood in terms of
an effective field theory setting, in
contrast to the physics yielding inflation.

String gas cosmology \cite{BV} (see also \cite{Perlt}, and see 
\cite{RHBSGrev, BWrev}  for a comprehensive review) is
a specific realization of the emergent universe paradigm.
It is a toy model of cosmology which results from coupling a
gas of fundamental strings to a background space-time metric.
The idea is to study consequences for cosmology of some of the
basic principles which distinguish string theory from point
particle theories. String theory contains new degrees of
freedom, and this leads to new symmetries. These new
degrees of freedom and new symmetries will lead to an
evolution of the early universe which is profoundly
different from what can be obtained if one works in the
restricted context of point particle theories only.

Let us be a bit more specific. First of all, it is assumed 
that the spatial sections are compact. For simplicity,
each spatial direction can be taken to be a circle with radius
$R$. Strings have three types
of states: {\it momentum modes} which represent the center
of mass motion of the string, {\it oscillatory modes} which
represent the fluctuations of the strings, and {\it winding
modes} counting the number of times a string wraps the 
spatial circle. Oscillatory and winding modes are not present
in point particle theories. Both lead to important
consequences in early universe cosmology.

\begin{figure} 
\includegraphics[height=6cm]{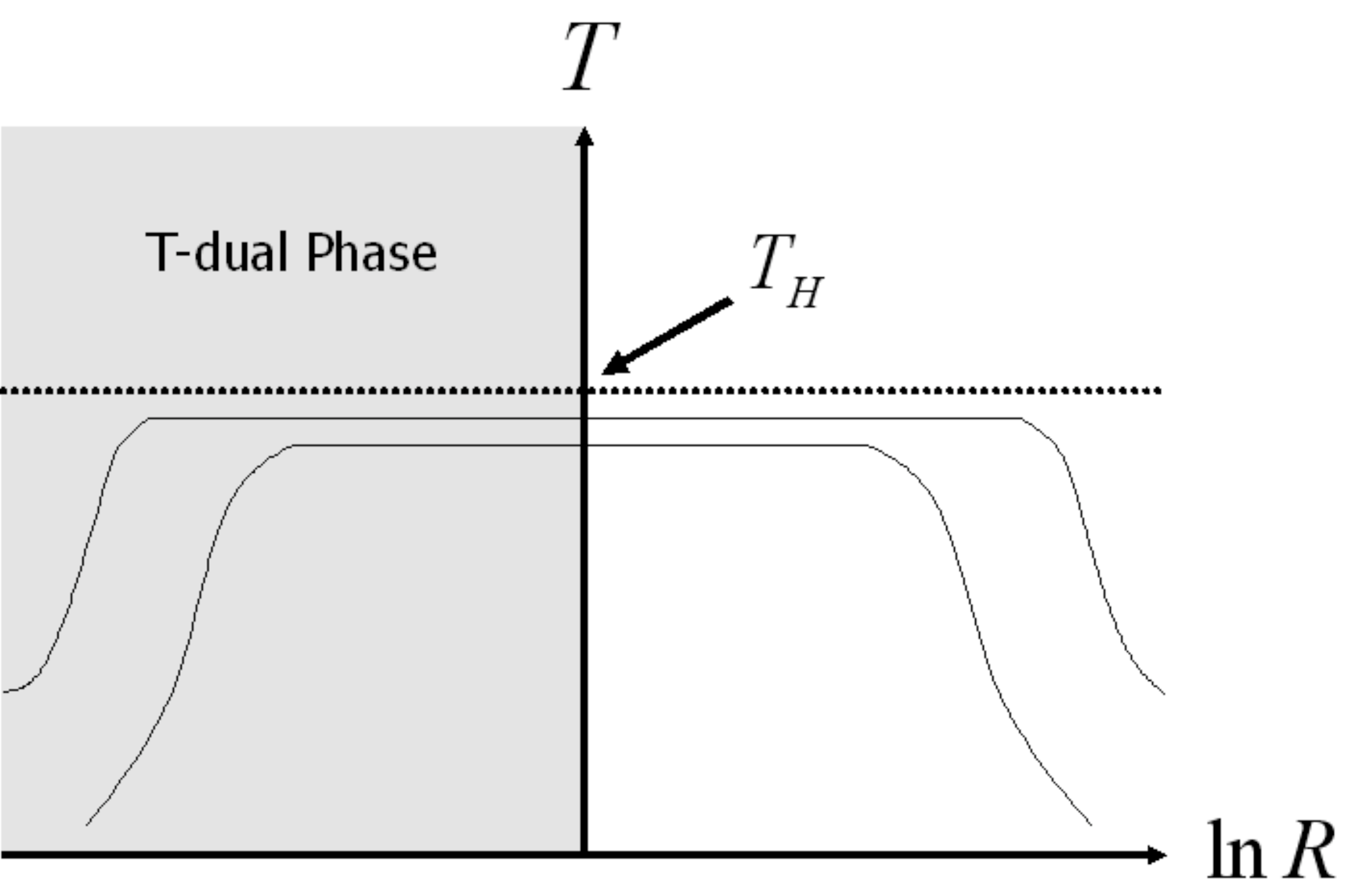}
\caption{The temperature (vertical axis) as a function of
radius (horizontal axis) of a gas of closed strings in thermal
equilibrium. Note the absence of a temperature singularity. The
range of values of $R$ for which the temperature is close to
the Hagedorn temperature $T_H$ depends on the total entropy
of the universe. The upper of the two curves corresponds to
a universe with larger entropy.}
\label{jirofig1}
\end{figure}

Since the number of string oscillatory states increases exponentially
with energy, there is a limiting  temperature for a gas of strings in
thermal equilibrium, the {\it Hagedorn temperature} \cite{Hagedorn}
$T_H$. Thus, if we take a box of strings and adiabatically decrease the box
size, the temperature will never diverge. This is the first indication that
string theory has the potential to resolve the cosmological singularity
problem. Figure (\ref{jirofig1}) is a sketch of how the temperature of a gas
of strings depends on the radius of space.

The second key feature of string theory upon which string gas cosmology
is based is {\it T-duality}. To introduce this symmetry, let us discuss the
radius dependence of the energy of the basic string states:
The energy of an oscillatory mode is independent of $R$, momentum
mode energies are quantized in units of $1/R$, i.e.
\be
E_n \, = \, n \mu \frac{{l_s}^2}{R} \, ,
\ee
where $l_s$ is the string length and $\mu$ is the mass per unit length of
a string. The winding mode energies are 
quantized in units of $R$, i.e.
\be
E_m \, = \, m \mu R \, ,
\ee
where both $n$ and $m$ are integers. Thus, a new symmetry of
the spectrum of string states emerges: Under the change
\be
R \, \rightarrow \, 1/R
\ee
in the radius of the torus (in units of  $l_s$)
the energy spectrum of string states is invariant if winding
and momentum quantum numbers are interchanged.
Perturbative string interactions  are consistent with this symmetry, and
thus T-duality is a symmetry of perturbative string theory. Postulating
that T-duality extends to non-perturbative string theory leads
\cite{Pol} to the need of adding D-branes to the list of fundamental
objects in string theory. With this addition, T-duality is expected
to be a symmetry of non-perturbative string theory
\footnote{Specifically, T-duality will take a spectrum of stable Type IIA branes
and map it into a corresponding spectrum of stable Type IIB branes
with identical masses \cite{Boehm}.}. As discussed in \cite{BV}, the 
above T-duality symmetry leads to
an equivalence between small and large spaces
 
Based on the above features of string theory, we can construct
the string gas realization of the emergent universe paradigm in
the following way: we assume that the universe starts in a quasi-static
phase during which the temperature of the string gas
hovers at the Hagedorn value \cite{Hagedorn}, the
maximal temperature of a gas of closed strings in
thermal equilibrium \footnote{The equations that govern the 
background of string gas cosmology
are not known. The Einstein equations are not the correct
equations since they do not obey the T-duality symmetry of
string theory. Many early studies of string gas cosmology were
based on using the dilaton gravity equations. However,
these equations are not satisfactory, either. Dilaton
gravity is a low energy approximation to the action of
string theory which only goes slightly beyond the Einstein
action. However, we expect that string theory correction 
terms to the low energy effective action of string theory will
be dominant in the Hagedorn phase, and that leading low
energy effective Lagrangians will give a very misleading
description of the dynamics.}. The string gas in this early
phase is dominated by strings winding the compact
spatial sections. The annihilation
of winding strings will produce string loops
\footnote{String loops have an equation of state like that
of radiation.} and lead
to a transition from the early quasi-static phase to the
radiation phase of Standard Cosmology.

As pointed out in \cite{BV}, the annihilation of winding
strings into string loops
is possible in at most three large spatial dimensions. This is
a simple dimension counting argument: string world sheets have
measure zero intersection probability in more than four large 
space-time dimensions. Hence, string gas cosmology
may provide a natural mechanism for explaining why there are
exactly three large spatial dimensions. Once three spatial
dimensions have started to expand, the string modes winding
other spatial sections cannot find eachother. Hence, the
radii of the extra spatial dimensions remain at the
string scale at all times. This argument was
supported by numerical studies of string evolution in three and
four spatial dimensions \cite{Mairi} (see also \cite{Cleaver})
\footnote{See \cite{Col2, Danos, Kabat} for some caveats to
this argument.}. The extra spatial dimensions are 
confined at fixed radius and shape by the string states
which have winding and momentum about them. Thus, string
gas cosmology provides a natural mechanism to stabilize
most of the string moduli which traditionally pose
a serious problem in any attempt to merge string theory
with cosmology. This issue is discussed in the
original papers \cite{Watson, Subodh1, Subodh2, Edna, Frey},
and the interested reader is referred to a review in
\cite{RHBSGrev}.

\section{Making Contact with Cosmological Observations}
\label{section:Obs}

Cosmology is currently in its golden ages in terms of
the wealth of new observational data which are being
collected every year. Thus, the main goal of modern
theoretical cosmology has become making contact between
early universe cosmology and the data about the structure
of the observed universe. The main tool being used is
the theory of cosmological perturbations. In the following
I give a brief overview of this theory and demonstrate
its application to both inflationary cosmology and to the
two alternative cosmological scenarios which were
discussed above. For a more technical overview, the
reader is referred in \cite{RHBrev}, and for an in-depth
survey to \cite{MFB}. 

\subsection{Cosmological Perturbations}

Cosmological perturbations are inhomogeneities in both
the metric of space-time and in the matter distribution.
To describe the generation and evolution of these
fluctuations, both General Relativity and
quantum mechanics are required - General Relativity
since the wavelengths of the inhomogeneities which
we are interested in are larger than the Hubble radius
\footnote{The Hubble radius is defined as $H^{-1}(t)$, where
$H(t)$ is the expansion rate of space at time $t$, and it
plays an important role in cosmology since it separates
length scales where matter interactions dominate (sub-Hubble)
from those larger ones where matter forces freeze out
and gravity dominates. It is important to realize that
the Hubble radius is not the same thing as the horizon, the
distance which light can travel starting at the initial
time. In all theories which have a chance of explaining the
origin of structure, the horizon must be much larger than
the Hubble radius.}
for a long period of time, and quantum mechanics since
in many models the origin of the fluctuations is
quantum mechanical \footnote{As discussed in the following
subsection, classical fluctuations are red-shifted during
inflation, hence leaving behind a vacuum state of matter. Thus,
the source of fluctuations should be from the quantum
vacuum. However, if there is a mechanism which generates
matter during inflation (like e.g. in the ``warm
inflation'' scenario \cite{Berera}), or if trans-Planckian
effects re-populate high energy modes, then classical
(e.g. thermal) fluctuations will dominate. In some alternatives
to inflation (e.g. the ``matter bounce'' scenario) it is
postulated that the origin of the inhomogeneities in quantum
mechanical, while in others (e.g. ``string gas cosmology'')
it is thermal.} . What makes the theory of
cosmological perturbations tractable is that the amplitude
of the fractional fluctuations is small today and hence (since gravity
is a purely attractive force) that it was even smaller in the
early universe. This justifies the linear analysis of the
generation and evolution of fluctuations.

In the context of a Universe with an inflationary period, 
the quantum origin of cosmological fluctuations
was first discussed in \cite{Mukh}  (see also \cite{Press,Sato} for
earlier ideas, and \cite{Starob0} for the corresponding
analysis of gravitational wave production). 
In particular, Mukhanov \cite{Mukh} and Press 
\cite{Press} realized that
in an exponentially expanding background, the curvature fluctuations
would be scale-invariant, and Mukhanov provided a quantitative
calculation which also yielded a logarithmic deviation from
exact scale-invariance. 

The basic idea of the theory of cosmological perturbations is
simple. In order to obtain the action for linearized cosmological
perturbations, we expand the action for gravity and matter to quadratic order in
the fluctuating degrees of freedom. The linear terms cancel
because the background is taken to satisfy the background
equations of motion. 

At first sight, it appears that there are ten degrees of freedom
for the metric fluctuations, in addition to the matter perturbations.
However, four of these degrees of freedom are equivalent to
space-time diffeomorphisms. To study the remaining six
degrees of freedom for metric fluctuations it proves very
useful to classify them according to how they transform under
spatial rotations. There are two scalar modes, two vector
modes and two tensor modes (which are the two helicity states
of gravitational waves). At linear order in cosmological
perturbation theory, scalar, vector and tensor modes decouple.
It is the scalar fluctuations which are the most important
since they describe how small matter perturbations lead to
fluctuations which grow in time and can become the very
inhomogeneous distribution of galaxies today.
For simple forms of matter such as scalar fields or perfect
fluids, the matter fluctuations couple only to the scalar
metric modes. These are the so-called ``cosmological
perturbations" which we focus on. 

If matter has no anisotropic stress, then one of the scalar
metric degrees of freedom disappears. In addition, one
of the Einstein constraint equations couples the remaining
metric degree of freedom to matter. Thus, if there is only
one matter component (e.g. one scalar matter field), there
is only one independent scalar cosmological fluctuation mode. 

To obtain the action and equation of motion for this mode,
we begin with the Einstein-Hilbert action for gravity and the
action for matter (which we take for simplicity to be a
scalar field $\varphi$ - for the more complicated
case of general hydrodynamical fluctuations the reader is
referred to \cite{MFB})
\begin{equation} \label{action}
S \, = \,  \int d^4x \sqrt{-g} \bigl[ - {1 \over {16 \pi G}} R
+ {1 \over 2} \partial_{\mu} \varphi \partial^{\mu} \varphi - V(\varphi)
\bigr] \, ,
\end{equation}
where $R$ is the Ricci curvature scalar.

The simplest way to proceed is to work in 
longitudinal gauge, in which the metric and matter take the form
(assuming no anisotropic stress)
\begin{eqnarray} \label{long}
ds^2 \, &=& \, a^2(\eta)\bigl[(1 + 2 \Phi(\eta, {\bf x}))d\eta^2
- (1 - 2 \Phi(t, {\bf x})) d{\bf x}^2 \bigr] \nonumber \\
\varphi(\eta, {\bf x}) \, 
&=& \, \varphi_0(\eta) + \delta \varphi(\eta, {\bf x}) \, ,
\end{eqnarray}
where $\eta$ in conformal time related to the physical time $t$
via $d \eta = a(t)^{-1} dt$. The two fluctuation variables
$\Phi$ and $\delta \varphi$ must be linked by the Einstein constraint
equations since there cannot be matter fluctuations without induced
metric fluctuations. 

The two nontrivial tasks of the lengthy \cite{MFB} computation 
of the quadratic piece of the action is to find
out what combination of $\delta \varphi$ and $\Phi$ gives the variable $v$
in terms of which the action has canonical kinetic term, and what the form
of the time-dependent mass is. This calculation involves inserting
the ansatz (\ref{long}) into the action (\ref{action}),
expanding the result to second order in the fluctuating fields, making
use of the background and of the constraint equations, and dropping
total derivative terms from the action. In the context of
scalar field matter, the quantum theory of cosmological
fluctuations was developed by Mukhanov \cite{Mukh2,Mukh3} and
Sasaki \cite{Sasaki}. The result is the following
contribution $S^{(2)}$ to the action quadratic in the
perturbations:
\begin{equation} \label{pertact}
S^{(2)} \, = \, {1 \over 2} \int d^4x \bigl[v'^2 - v_{,i} v_{,i} + 
{{z''} \over z} v^2 \bigr] \, ,
\end{equation}
where the canonical variable $v$ (the ``Sasaki-Mukhanov variable'' introduced
in \cite{Mukh3}) is given by
\begin{equation} \label{Mukhvar}
v \, = \, a \bigl[ \delta \varphi + {{\varphi_0^{'}} \over {\cal H}} \Phi
\bigr] \, ,
\end{equation}
with ${\cal H} = a' / a$, and where
\begin{equation} \label{zvar}
z \, = \, {{a \varphi_0^{'}} \over {\cal H}} \, .
\end{equation}

As long as
the equation of state does not change over time
\begin{equation} \label{zaprop}
z(\eta) \, \sim \, a(\eta) \, .
\end{equation}
Note that the variable $v$ is related to the curvature
perturbation ${\cal R}$ in comoving coordinates introduced
in \cite{Lyth0} and closely related to the variable $\zeta$ used
in \cite{BST}:
\begin{equation} \label{Rvar}
v \, = \, z {\cal R} \, .
\end{equation}

The equation of motion which follows from the action (\ref{pertact}) is
(in momentum space)
\begin{equation} \label{pertEOM2}
v_k^{''} + k^2 v_k - {{z^{''}} \over z} v_k \, = \, 0 \, ,
\end{equation}
where $v_k$ is the k'th Fourier mode of $v$. 
The mass term in the above equation is in general
given by the Hubble scale (the scale whose wave-number will
be denoted $k_H$). Thus, it immediately follows that on small
length scales, i.e. for
$k > k_H$, the solutions for $v_k$ are constant amplitude oscillations . 
These oscillations freeze out at Hubble radius crossing,
i.e. when $k = k_H$. On longer scales ($k \ll k_H$), there is
a mode of  $v_k$ which scales as $z$. This mode is the dominant
one in an expanding universe, but not in a contracting one.

Given the action (\ref{pertact}), the cosmological
perturbations can be quantized by canonical quantization (in the same
way that a scalar matter field on a fixed cosmological background
is quantized \cite{BD}). 

The final step in the quantum theory of cosmological perturbations
is to specify an initial state. For an initial vacuum state we
have harmonic oscillator ground state initial conditions for each
Fourier mode of the field:
\begin{eqnarray} \label{incond}
v_k(\eta_i) \, = \, {1 \over {\sqrt{2 k}}} \\
v_k^{'}(\eta_i) \, = \, {{\sqrt{k}} \over {\sqrt{2}}} \, \, \nonumber
\end{eqnarray} 
where $\eta_i$ is the conformal time corresponding
to the initial physical time $t_i$.

In an expanding background, the scaling 
$v_k \, \sim \, z \, \sim \, a $
implies that the wave function of the quantum variable $v_k$ which
performs quantum vacuum fluctuations on sub-Hubble scales,
stops oscillating on super-Hubble scales and instead is
squeezed (the amplitude increases in configuration space
but decreases in momentum space). This squeezing corresponds
to quantum particle production, and it is one of the two
conditions which are required for the classicalization of
the fluctuations \footnote{These conditions are necessary, but
there is some debate as to whether they are sufficient to solve
the ``cosmic measurement problem" (see e.g. \cite{Sudarsky}
for an analysis from the physics perspective, and \cite{HZ2}
for a discussion in the framework of philosophy of physics).} 
The second condition is decoherence which
is induced by the non-linearities in the dynamical system
which are inevitable since the Einstein action leads to
highly nonlinear equations (see \cite{Starob3} for an in-depth
discussion of this point, and \cite{Martineau} for related
work). Note that the squeezing of cosmological fluctuations on
super-Hubble scales occurs in all models, in particular
in string gas cosmology and in the bouncing universe
scenario since also in these scenarios perturbations
propagate on super-Hubble scales for a long period of
time. In a contracting phase, the dominant
mode of $v_k$ on super-Hubble scales is also growing
(the mode $v_k \sim a$ is now a decaying
one). Thus, the squeezing of fluctuations on super-Hubble
scales and the resulting classicalization of the perturbations
also occurs. 

\subsection{Fluctuations in Inflationary Cosmology}

In inflationary cosmology (see Fig. 1), we pick some initial
time $t_i$ after the beginning of the inflationary
phase and start with quantum vacuum initial conditions
\be \label{IC}
v_k(\eta_i) \, = \, \frac{1}{\sqrt{2 k}} \, 
\ee
for all values of $k$ for which the wavelength is smaller than
the Hubble radius at the initial time $t_i$. Since $v_k$
oscillates on sub-Hubble scales, 
the amplitude remains unchanged until the mode
exits the Hubble radius at the time $t_H(k)$ given by
\begin{equation} \label{Hubble3}
a^{-1}(t_H(k)) k \, = \, H \, .
\end{equation}

We need to compute the power spectrum ${\cal P}_{\cal R}(k)$ 
of the curvature fluctuation ${\cal R}$ defined in (\ref{Rvar}) at
some late time $t$ when the modes are super-Hubble.
We first relate the power spectrum via the growth $v_k \sim a$
on super-Hubble scales to the power spectrum at the time 
$t_H(k)$ and then use the constancy of the amplitude
of $v$ on sub-Hubble scales to relate it to the initial conditions
(\ref{IC}). Thus
\begin{eqnarray} \label{finalspec1}
{\cal P}_{\cal R}(k, t) \, \equiv  \, k^3 {\cal R}_k^2(t) \, 
&=& \, k^3 z^{-2}(t) |v_k(t)|^2 \\
&=& \, k^3 z^{-2}(t) \bigl( {{a(t)} \over {a(t_H(k))}} \bigr)^2
|v_k(t_H(k))|^2 \nonumber \\
&=& \, k^3 z^{-2}(t_H(k)) |v_k(t_H(k))|^2 \nonumber \\
&\sim& \, k^3 \bigl( \frac{a(t)}{z(t)} \bigr)^2 a^{-2}(t_H(k)) |v_k(t_i)|^2 \, , \nonumber
\end{eqnarray}
where in the final step we have used (\ref{zaprop}) and the
constancy of the amplitude of $v$ on sub-Hubble scales. 
Making use of the condition (\ref{Hubble3}) 
for Hubble radius crossing, and of the
initial conditions (\ref{IC}), we immediately see that
\begin{equation} \label{finalspec2}
{\cal P}_{\cal R}(k, t) \, \sim \, \bigl( \frac{a(t)}{z(t)} \bigr)^2 
k^3 k^{-2} k^{-1} H^2 \, ,
\end{equation}
and that thus a scale invariant \footnote{Scale invariance means
that the power spectrum is independent of $k$.} power spectrum results.
Since $H$ decreases slowly as inflation proceeds, the amplitude of
the spectrum is slightly smaller for short wavelengths (which exit the
Hubble radius later) than for longer ones. Thus, the spectrum
has a slight tilt. Note that the alternative models which we discuss
below also predict a slight red tilt of the spectrum. Current
precision observations indicate that there indeed is a slight red
tilt.

\subsection{Fluctuations in the Matter Bounce}

In the Matter Bounce paradigm (see Fig. 4) the universe starts out large
but cold. Hence, it is reasonable to assume that fluctuations
on cosmological scales begin in their vacuum state
\footnote{The vacuum state of the fluctuation field $v$. This prescription
does not require matter to be a quantum field. Cosmological
perturbations can be quantized \cite{MFB} even if matter
is a usual hydrodynamical fluid.}. The vacuum
spectrum is blue, i.e. there is more power on short wavelengths.
To obtain a scale-invariant spectrum, a mechanism is required
which boosts long wavelength modes relative to short wavelength
ones. In a contracting universe long wavelength modes exit
the Hubble radius earlier than short wavelength modes. If
the fluctuation modes grow on super-Hubble scales, then we
have a mechanism which boosts the long wavelength fluctuations.

In a contracting universe the mode $v_k \sim a$ is decaying
and unimportant. The second mode, 
which is sub-dominant in an expanding universe,
becomes the dominant one, and grows in time 
with a rate which depends on the equation of state of
the background. It turns out that the growth rate in the
case of a matter-dominated phase of contraction is exactly
the correct one to turn an initial vacuum spectrum into
a scale-invariant one, as was realized in \cite{Wands, Fabio2}.

The analysis is not very complicated: if the contracting phase 
is matter-dominated then $a(t) \sim t^{2/3}$, $\eta(t) \sim t^{1/3}$ 
and the dominant mode of $v$ scales as $\eta^{-1}$. Thus,
the power spectrum of curvature fluctuations becomes
\bea
P_{{\cal R}}(k, \eta) \, &\sim& k^3 |v_k(\eta)|^2 a^{-2}(\eta) \\
&\sim& \, k^3 |v_k(\eta_H(k))|^2 \bigl( \frac{\eta_H(k)}{\eta} \bigr)^2 \, 
\sim \, k^{3 - 1 - 2} \nonumber \\
&\sim& \, {\rm const}  \, , \nonumber 
\eea
where we have made use of
the assumption that we have a vacuum spectrum at Hubble radius crossing.

At this point we have shown that the spectrum of fluctuations is
scale-invariant on super-Hubble scales before the bounce phase.
The evolution during the bounce depends in principle on the
specific realization of the non-singular bounce. In any concrete
model, the equations of motion can be solved 
numerically without approximation during the bounce.
Alternatively, we can solve them approximately using analytical techniques.
The bottom line of a large number of studies in
specific realizations of non-singular bouncing cosmologies
(see e.g. \cite{ABB, Saremi, Cai1, Chunshan, HLbounce2, BKPPT} 
is that on length scales large compared to the time of the bounce, the
spectrum of curvature fluctuations is not changed during the
bounce phase. Since typically the bounce time is set by
a microphysical scale whereas the wavelength of fluctuations
which we observe today is macroscopic (about $1 {\rm mm}$
if the bounce scale is set by the particle physics GUT scale),
we conclude that for scales relevant to current observations
the spectrum is unchanged during the bounce. This completes
the demonstration that a non-singular matter bounce leads
to a scale-invariant spectrum of cosmological perturbations
after the bounce provided that the initial spectrum on
sub-Hubble scales is vacuum. 

A specific signal of growing curvature fluctuations on super-Hubble
scales in the contracting matter-dominated phase is a special
form \cite{bispectrum} of the ``bispectrum'', the three point correlation function
of the CMB. The predicted amplitude of the bispectrum is also
significantly larger than what is obtained in simple inflationary
models. Thus, the way to observationally distinguish between the
matter bounce and the inflationary paradigms is via precision
observations of the bispectrum.

I close this section with a side comment which might be of
interest to philosophers of cosmology: the fact that 
fluctuations grow both in the contracting and
expanding phase has implications for attempts to build a
cyclic cosmology. In four space-time dimensions 
it is impossible to construct a model which is cyclic in
the presence of fluctuations  - the
growth of fluctuations breaks any cyclicity which may
be present at the background level. 
Four space-time-dimensional cyclic background cosmologies
are not predictive - the index of the power spectrum changes
from cycle to cycle \cite{processing}. Note that the
cyclic version of the Ekpyrotic scenario \cite{Ekpcyclic}
avoids these problems because it is not cyclic in the
above sense: it is a higher space-time-dimensional model
in which the radius of an extra dimension evolves
cyclically, but the four-dimensional scale factor does
not.

\subsection{Fluctuations in String Gas Cosmology}

In String Gas Cosmology (see Fig. 6) the physical 
wavelength of a fluctuation mode is constant in the Hagedorn phase
since space is static. The Hubble radius is infinite in
this phase. Hence, as in the case of inflationary cosmology,
fluctuation modes begin sub-Hubble during the Hagedorn
phase, and thus a causal generation mechanism
for fluctuations is possible.

However, the physics of the generation mechanism is
very different. In the case of inflationary cosmology,
fluctuations are assumed to start as quantum vacuum
perturbations because classical inhomogeneities are
red-shifting. In contrast, in the Hagedorn phase of string gas
cosmology there is no red-shifting of classical matter.
Hence, it is the fluctuations in the classical matter which
dominate. Since classical matter is a string gas, the
dominant fluctuations are string thermodynamic fluctuations. 
It turns out that the holographic scaling of the specific
heat capacity of a gas of closed strings is key to obtaining
a scale-invariant spectrum of curvature fluctuations.

The proposal for string gas structure formation is the following 
\cite{NBV} (see \cite{BNPV2} for a more detailed description).
For a fixed co-moving scale with wavenumber $k$ we compute the matter
fluctuations of the thermal string gas,
while the scale is sub-Hubble (and therefore gravitational
effects are sub-dominant). When the scale exits the Hubble radius
at time $t_i(k)$ we use the gravitational constraint equations to
determine the induced metric fluctuations, which are then propagated
to late times using the usual equations of gravitational perturbation
theory. Since the scales we are interested
in are in the far infrared, we use the Einstein constraint equations for
fluctuations.

Assuming that the fluctuations are described by the perturbed Einstein
equations (they are {\it not} if the dilaton is not fixed 
\cite{Betal,KKLM}), then the spectra of cosmological perturbations
$\Phi$ and gravitational waves $h$ are given by the energy-momentum 
fluctuations in the following way \cite{BNPV2}
\be \label{scalarexp} 
\langle|\Phi(k)|^2\rangle \, = \, 16 \pi^2 G^2 
k^{-4} \langle\delta T^0{}_0(k) \delta T^0{}_0(k)\rangle \, , 
\ee 
where the pointed brackets indicate expectation values, and 
\be 
\label{tensorexp} \langle|h(k)|^2\rangle \, = \, 16 \pi^2 G^2 
k^{-4} \langle\delta T^i{}_j(k) \delta T^i{}_j(k)\rangle \,, 
\ee 
where on the right hand side of (\ref{tensorexp}) we mean the 
average over the correlation functions with $i \neq j$, and
$h$ is the amplitude of the gravitational waves \footnote{The
gravitational wave tensor $h_{i j}$ can be written as the
amplitude $h$ multiplied by a constant polarization tensor.}.
 
The equation (\ref{scalarexp}) is used to determine the spectrum of
scalar metric fluctuations. We first calculate the 
root mean square energy density fluctuations in a sphere of
radius $R = k^{-1}$. For a system in thermal equilibrium they 
are given by the specific heat capacity $C_V$ via 
\be \label{cor1b}
\langle \delta\rho^2 \rangle \,  = \,  \frac{T^2}{R^6} C_V \, . 
\ee 
The specific  heat of a gas of closed strings
on a torus of radius $R$ can be derived from the partition
function of a gas of closed strings. This computation was
carried out in \cite{Deo} (see also \cite{Ali}) with the result
\be \label{specheat2b} 
C_V  \, \approx \, 2 \frac{R^2/\ell^3}{T \left(1 - T/T_H\right)}\, . 
\ee 
The specific heat capacity scales holographically with the size
of the box. This result follows rigorously from evaluating the
string partition function in the Hagedorn phase. The result, however,
can also be understood heuristically: in the Hagedorn phase the
string winding modes are crucial. These modes look like point
particles in one less spatial dimension. Hence, we expect the
specific heat capacity to scale like in the case of point particles
in one less dimension of space \footnote{We emphasize that it was
important for us to have compact spatial dimensions in order to
obtain the winding modes which are crucial to get the holographic
scaling of the thermodynamic quantities.}.

With these results, the power spectrum $P(k)$ of scalar metric fluctuations can
be evaluated as follows
\bea \label{power2} 
P_{\Phi}(k) \, & \equiv & \, {1 \over {2 \pi^2}} k^3 |\Phi(k)|^2 \\
&=& \, 8 G^2 k^{-1} <|\delta \rho(k)|^2> \, . \nonumber \\
&=& \, 8 G^2 k^2 <(\delta M)^2>_R \nonumber \\ 
               &=& \, 8 G^2 k^{-4} <(\delta \rho)^2>_R \nonumber \\
&=& \, 8 G^2 {T \over {\ell_s^3}} {1 \over {1 - T/T_H}} 
\, , \nonumber 
\eea 
where in the first step we have used (\ref{scalarexp}) to replace the 
expectation value of $|\Phi(k)|^2$ in terms of the correlation function 
of the energy density, and in the second step we have made the 
transition to position space. 

The first conclusion from the result (\ref{power2}) is that the spectrum
is approximately scale-invariant ($P(k)$ is independent of $k$). It is
the `holographic' scaling $C_V(R) \sim R^2$ which is responsible for the
overall scale-invariance of the spectrum of cosmological perturbations.
However, there is a small $k$ dependence which comes from the fact
that in the above equation for a scale $k$ 
the temperature $T$ is to be evaluated at the
time $t_i(k)$. Thus, the factor $(1 - T/T_H)$ in the 
denominator is responsible 
for giving the spectrum a slight dependence on $k$. Since
the temperature slightly decreases as time increases at the
end of the Hagedorn phase, shorter wavelengths for which
$t_i(k)$ occurs later obtain a smaller amplitude. Thus, the
spectrum has a slight red tilt.

As discovered in \cite{BNPV1}, the spectrum of gravitational
waves is also nearly scale invariant. However, in the expression
for the spectrum of gravitational waves the factor $(1 - T/T_H)$
appears in the numerator, thus leading to a slight blue tilt in
the spectrum. This is a prediction with which the cosmological
effects of string gas cosmology can be distinguished from those
of inflationary cosmology, where quite generically a slight red
tilt for gravitational waves results. The physical reason for the
blue tilt in string gas cosmology is that
large scales exit the Hubble radius earlier when the pressure
and hence also the off-diagonal spatial components of $T_{\mu \nu}$
are closer to zero. For more details the interested reader is
referred to the original literature.

\section{Discussion}
\label{section:Dis}

As I hope to have convinced the reader, inflationary cosmology,
in spite of its spectacular phenomenological success, suffers
from several conceptual problems and should at this point not
yet be considered as an established theory of the early universe.
The recent speculations concerning the ``inflationary multiverse'' 
face even more serious problems and should be viewed with
caution.

I have presented two alternative scenarios which are in
equally good agreement with current observations on the distribution
of galaxies and anisotropies of the CMB as the inflationary
paradigm
\footnote{At this point it must be emphasized again that obtaining
a scale-invariant spectrum of curvature fluctuations is
a very non-trivial requirement for early universe cosmology.
For example, initial vacuum fluctuations generated in a
contracting universe are not scale-invariant unless the
relevant phase of matter contraction is dominated by
non-relativistic matter.} These two scenarios make predictions for future
observations which are different from those of inflation.
Which of these scenarios is actually realized in nature
will ultimately be determined by observations.

The two alternative scenarios which I focused on are the
{\it Matter Bounce} paradigm involving a non-singular
background cosmology which starts in a matter-dominated
phase of contraction, and the {\it Emergent Universe}
picture as realized in {\it String Gas Cosmology}.
These are two scenarios the author has worked on. However,
there are other scenarios, a particularly promising one
being the {\it Ekpyrotic scenario} \cite{Ekp}, which also
leads to a picture in which time runs from $- \infty$ to
$+ \infty$.

None of the alternative scenarios are without problems.
In fact, one may argue that none of them address all of
the classic problems of Standard Big Bang cosmology as well
as inflation does. The matter bounce scenario suffers
from an instability to the development of anisotropies,
and emergent scenarios do not as nicely explain the large
size and entropy of the universe as inflation does. However,
both of them are free from one of the key problems of
inflationary cosmology, namely the {\it trans-Planckian
problem} for cosmological perturbations.


{\it Acknowledgements}: I wish to thank Henrik Zinkernagel for the invitation to participate
and lecture at this stimulating workshop, and also for useful comments
on the draft of these lecture notes. I wish to thank all of my 
collaborators on whose
work I have drawn. This work has been supported in part by funds from
an NSERC Discovery Grant and from the Canada Research Chair
program. I also acknowledge support from the Killam Foundation
for the period 9/09 - 8/11.



\begin{thebibliography}{00}



\bibitem{Guth}
Guth AH, 
 ``The Inflationary Universe: A Possible Solution To The Horizon And Flatness
 Problems,''
  Phys.\ Rev.\  D {\bf 23}, 347 (1981).
  
\bibitem{Brout}
R.~Brout, F.~Englert and E.~Gunzig,
  ``The Creation Of The Universe As A Quantum Phenomenon,''
  Annals Phys.\  {\bf 115}, 78 (1978).
  
\bibitem{Starob}
A.~A.~Starobinsky,
  ``A New Type Of Isotropic Cosmological Models Without Singularity,''
  Phys.\ Lett.\ B {\bf 91}, 99 (1980).
  
\bibitem{Sato}
K.~Sato,
  ``First Order Phase Transition Of A Vacuum And Expansion Of The Universe,''
  Mon.\ Not.\ Roy.\ Astron.\ Soc.\  {\bf 195}, 467 (1981).

\bibitem{Mukh}
V. Mukhanov and G. Chibisov,
  ``Quantum Fluctuation And Nonsingular Universe. (In Russian),''
  JETP Lett.\  {\bf 33}, 532 (1981)
  [Pisma Zh.\ Eksp.\ Teor.\ Fiz.\  {\bf 33}, 549 (1981)].

\bibitem{Press}
W. Press,
``Spontaneous production of the Zel'dovich spectrum of cosmological 
fluctuations'',
 Phys. Scr. {\bf 21}, 702 (1980).

\bibitem{WMAP}
C.~L.~Bennett {\it et al.},
   ``First Year Wilkinson Microwave Anisotropy Probe (WMAP) Observations:
  Preliminary Maps and Basic Results,''
  Astrophys.\ J.\ Suppl.\  {\bf 148}, 1 (2003)
  [arXiv:astro-ph/0302207].

\bibitem{Rashid}
 R.~A.~Sunyaev, Y.~.B.~Zeldovich,
  ``Small scale fluctuations of relic radiation,''
  Astrophys.\ Space Sci.\  {\bf 7}, 3-19 (1970).

\bibitem{Peebles}
  P.~J.~E.~Peebles, J.~T.~Yu,
  ``Primeval adiabatic perturbation in an expanding universe,''
  Astrophys.\ J.\  {\bf 162}, 815-836 (1970).

\bibitem{COBE}
G.~F.~Smoot, C.~L.~Bennett, A.~Kogut, E.~L.~Wright, J.~Aymon, N.~W.~Boggess, E.~S.~Cheng and G.~De Amici {\it et al.},
  ``Structure in the COBE differential microwave radiometer first year maps,''
  Astrophys.\ J.\  {\bf 396}, L1 (1992).
  
\bibitem{Boomerang}
 P.~D.~Mauskopf {\it et al.}  [Boomerang Collaboration],
  ``Measurement of a peak in the cosmic microwave background power spectrum from the North American test flight of BOOMERANG,''
  Astrophys.\ J.\  {\bf 536}, L59 (2000)
  [astro-ph/9911444].
 
\bibitem{Linde}
A.~D.~Linde,
  ``Chaotic Inflation,''
  Phys.\ Lett.\ B {\bf 129}, 177 (1983).
  
\bibitem{attractor}
 R.~H.~Brandenberger and J.~H.~Kung,
  ``Chaotic Inflation As An Attractor In Initial Condition Space,''
  Phys.\ Rev.\ D {\bf 42}, 1008 (1990);\\
  R.~H.~Brandenberger, H.~Feldman and J.~Kung,
  ``Initial conditions for chaotic inflation,''
  Phys.\ Scripta T {\bf 36}, 64 (1991).
  
\bibitem{Shaposh}
 F.~L.~Bezrukov and M.~Shaposhnikov,
  ``The Standard Model Higgs boson as the inflaton,''
  Phys.\ Lett.\  B {\bf 659}, 703 (2008)
  [arXiv:0710.3755 [hep-th]].

\bibitem{inflrevs}
A.~D.~Linde,
  ``Inflationary Cosmology,''
  Lect.\ Notes Phys.\  {\bf 738}, 1 (2008)
  [arXiv:0705.0164 [hep-th]];\\
  A.~Mazumdar and J.~Rocher,
  ``Particle physics models of inflation and curvaton scenarios,''
  Phys.\ Rept.\  {\bf 497}, 85 (2011)
  [arXiv:1001.0993 [hep-ph]].
  
\bibitem{Starob2}
A.~A.~Starobinsky,
  ``Stochastic De Sitter (inflationary) Stage In The Early Universe,''
  In *De Vega, H.j. ( Ed.), Sanchez, N. ( Ed.): Field Theory, Quantum Gravity and Strings*, 107-126
 
\bibitem{eternal}
A.~Vilenkin,
  ``The Birth of Inflationary Universes,''
  Phys.\ Rev.\ D {\bf 27}, 2848 (1983);\\
A.~D.~Linde,
  ``Eternally Existing Selfreproducing Chaotic Inflationary Universe,''
  Phys.\ Lett.\ B {\bf 175}, 395 (1986).
  
\bibitem{number}
F.~Denef and M.~R.~Douglas,
  ``Distributions of flux vacua,''
  JHEP {\bf 0405}, 072 (2004)
  [hep-th/0404116].
  
\bibitem{multiverse}
 {\it Universe or Multiverse?}, B. Carr ed., Cambridge University Press (2007),
 
\bibitem{swamp}
H.~Ooguri and C.~Vafa,
  ``On the Geometry of the String Landscape and the Swampland,''
  Nucl.\ Phys.\ B {\bf 766}, 21 (2007)
  [hep-th/0605264].
  
\bibitem{back}
V.~F.~Mukhanov, L.~R.~W.~Abramo and R.~H.~Brandenberger,
  ``On the Back reaction problem for gravitational perturbations,''
  Phys.\ Rev.\ Lett.\  {\bf 78}, 1624 (1997)
  [gr-qc/9609026];\\
  R.~H.~Brandenberger,
  ``Back reaction of cosmological perturbations and the cosmological constant problem,''
  hep-th/0210165.
  
\bibitem{Adams}
 F.~C.~Adams, K.~Freese and A.~H.~Guth,
  ``Constraints on the scalar field potential in inflationary models,''
  Phys.\ Rev.\  D {\bf 43}, 965 (1991).

\bibitem{RHBrev3}
 R.~H.~Brandenberger,
  ``Inflationary cosmology: Progress and problems,''
  arXiv:hep-ph/9910410.

\bibitem{Shenker}
N.~Kaloper, M.~Kleban, A.~E.~Lawrence and S.~Shenker,
  ``Signatures of short distance physics in the cosmic microwave background,''
  Phys.\ Rev.\ D {\bf 66}, 123510 (2002)
  [hep-th/0201158].
  
\bibitem{Jerome1}
R.~H.~Brandenberger and J.~Martin, 
``The Robustness of inflation to changes in superPlanck scale physics,''
Mod.~Phys.~Lett.~A~{\bf 16}, 999
(2001), [arXiv:astro-ph/0005432];\\
J.~Martin and R.~H.~Brandenberger,
``The TransPlanckian problem of inflationary cosmology,''
Phys.~Rev.~D~{\bf 63}, 123501 (2001), [arXiv:hep-th/0005209].

\bibitem{Niemeyer}
 J.~C.~Niemeyer,
``Inflation with a high frequency cutoff,''
Phys.~Rev.~D~{\bf 63}, 123502 (2001),
[arXiv:astro-ph/0005533];\\
J.~C.~Niemeyer and R.~Parentani,
``Minimal modifications of the primordial power spectrum from an  adiabatic
  short distance cutoff,''
  Phys.~Rev.~D~{\bf 64}, 101301 (2001),
[arXiv:astro-ph/0101451];\\
S.~Shankaranarayanan,
  ``Is there an imprint of Planck scale physics on inflationary cosmology?,''
  Class.\ Quant.\ Grav.\  {\bf 20}, 75 (2003)
  [arXiv:gr-qc/0203060].

\bibitem{Zhang}
R.~Brandenberger and X.~-m.~Zhang,
  ``The Trans-Planckian Problem for Inflationary Cosmology Revisited,''
  arXiv:0903.2065 [hep-th].
  
\bibitem{HE}
S.~W.~Hawking and G.~F.~R.~Ellis,
  ``The Large scale structure of space-time,''
{\it  Cambridge University Press, Cambridge, 1973}

\bibitem{Borde}
 A.~Borde and A.~Vilenkin,
  ``Eternal inflation and the initial singularity,''
  Phys.\ Rev.\ Lett.\  {\bf 72}, 3305 (1994)
  [arXiv:gr-qc/9312022].

\bibitem{HZ1}
H. Zinkernagel,
``Did Time Have a Beginning?",
Int. Studies in the Philosophy of Science {\bf 22}, 237 (2008).

\bibitem{Novello}
M.~Novello and S.~E.~P.~Bergliaffa,
  ``Bouncing Cosmologies,''
  Phys.\ Rept.\  {\bf 463}, 127 (2008)
  [arXiv:0802.1634 [astro-ph]].
  
\bibitem{PBB}
M.~Gasperini and G.~Veneziano,
  ``Pre - big bang in string cosmology,''
  Astropart.\ Phys.\  {\bf 1}, 317 (1993)
  [arXiv:hep-th/9211021].
  
\bibitem{BV}
R.~H.~Brandenberger and C.~Vafa,
  ``Superstrings in the Early Universe,''
  Nucl.\ Phys.\  B {\bf 316}, 391 (1989).
  
\bibitem{Ekp}
J.~Khoury, B.~A.~Ovrut, P.~J.~Steinhardt and N.~Turok,
  ``The ekpyrotic universe: Colliding branes and the origin of the hot big
  bang,''
  Phys.\ Rev.\  D {\bf 64}, 123522 (2001)
  [arXiv:hep-th/0103239].

\bibitem{Ekpcyclic}
P.~J.~Steinhardt and N.~Turok,
  ``Cosmic evolution in a cyclic universe,''
  Phys.\ Rev.\ D {\bf 65}, 126003 (2002)
  [hep-th/0111098].

\bibitem{Thorsten}
T.~J.~Battefeld, S.~P.~Patil and R.~Brandenberger,
  ``Non-singular perturbations in a bouncing brane model,''
  Phys.\ Rev.\ D {\bf 70}, 066006 (2004)
  [hep-th/0401010].
  
\bibitem{Wands}
D.~Wands,
  ``Duality invariance of cosmological perturbation spectra,''
  Phys.\ Rev.\  D {\bf 60}, 023507 (1999)
  [arXiv:gr-qc/9809062].
  
\bibitem{Fabio2}
F.~Finelli and R.~Brandenberger,
  ``On the generation of a scale-invariant spectrum of adiabatic  fluctuations
  in cosmological models with a contracting phase,''
  Phys.\ Rev.\  D {\bf 65}, 103522 (2002)
  [arXiv:hep-th/0112249].
  
\bibitem{KPT}
C.~Kounnas, H.~Partouche and N.~Toumbas,
  ``Thermal duality and non-singular cosmology in d-dimensional superstrings,''
 Nucl.\ Phys.\  B {\bf 855}, 280 (2012)
  [arXiv:1106.0946 [hep-th]];\\
C.~Kounnas, H.~Partouche and N.~Toumbas,
  ``S-brane to thermal non-singular string cosmology,''
  arXiv:1111.5816 [hep-th].

\bibitem{BKPPT}
R. Brandenberger, C. Kounnas, H. Partouche, S. Patil and N. Toumbas,
``Fluctuations in Non-Singular Bouncing Cosmologies from Type II Superstrings",
to be submitted.

\bibitem{quintom}
B.~Feng, X.~L.~Wang and X.~M.~Zhang,
  ``Dark Energy Constraints from the Cosmic Age and Supernova,''
  Phys.\ Lett.\  B {\bf 607}, 35 (2005)
  [arXiv:astro-ph/0404224];\\
B.~Feng, M.~Li, Y.~S.~Piao and X.~Zhang,
  ``Oscillating quintom and the recurrent universe,''
  Phys.\ Lett.\  B {\bf 634}, 101 (2006)
  [arXiv:astro-ph/0407432];\\
Y.~F.~Cai, T.~Qiu, Y.~S.~Piao, M.~Li and X.~Zhang,
  ``Bouncing Universe with Quintom Matter,''
  JHEP {\bf 0710}, 071 (2007)
  [arXiv:0704.1090 [gr-qc]];\\
 Y.~F.~Cai, T.~T.~Qiu, J.~Q.~Xia and X.~Zhang,
  ``A Model Of Inflationary Cosmology Without Singularity,''
  Phys.\ Rev.\  D {\bf 79}, 021303 (2009)
  [arXiv:0808.0819 [astro-ph]].

\bibitem{Cai1}
Y.~F.~Cai, T.~Qiu, R.~Brandenberger, Y.~S.~Piao and X.~Zhang,
  ``On Perturbations of Quintom Bounce,''
  JCAP {\bf 0803}, 013 (2008)
  [arXiv:0711.2187 [hep-th]];\\
    Y.~F.~Cai and X.~Zhang,
  ``Evolution of Metric Perturbations in Quintom Bounce model,''
JCAP {\bf 0906}, 003 (2009)
  [arXiv:0808.2551 [astro-ph]];\\
Y.~F.~Cai, T.~Qiu, R.~Brandenberger and X.~Zhang,
  ``A Nonsingular Cosmology with a Scale-Invariant Spectrum of Cosmological
  Perturbations from Lee-Wick Theory,''
Phys.\ Rev.\  D {\bf 80}, 023511 (2009)
  [arXiv:0810.4677 [hep-th]].
  
\bibitem{RHBbouncerev}
R.~H.~Brandenberger,
  ``Alternatives to Cosmological Inflation,''
  arXiv:0902.4731 [hep-th];\\
R.~H.~Brandenberger,
  ``Cosmology of the Very Early Universe,''
  AIP Conf.\ Proc.\  {\bf 1268}, 3-70 (2010).
  [arXiv:1003.1745 [hep-th]];\\
R.~H.~Brandenberger,
  ``Introduction to Early Universe Cosmology,''
  PoS ICFI {\bf 2010}, 001 (2010)
  [arXiv:1103.2271 [astro-ph.CO]].

\bibitem{ghost}
J.~M.~Cline, S.~Jeon and G.~D.~Moore,
  ``The phantom menaced: Constraints on low-energy effective ghosts,''
  Phys.\ Rev.\  D {\bf 70}, 043543 (2004)
  [arXiv:hep-ph/0311312].

\bibitem{Creminelli}
P.~Creminelli and L.~Senatore,
  ``A smooth bouncing cosmology with scale invariant spectrum,''
  JCAP {\bf 0711}, 010 (2007)
  [arXiv:hep-th/0702165];\\
E.~I.~Buchbinder, J.~Khoury and B.~A.~Ovrut,
  ``New Ekpyrotic Cosmology,''
  Phys.\ Rev.\  D {\bf 76}, 123503 (2007)
  [arXiv:hep-th/0702154].
  
\bibitem{Chunshan}
 C.~Lin, R.~H.~Brandenberger and L.~P.~Levasseur,
  ``A Matter Bounce By Means of Ghost Condensation,''
  JCAP {\bf 1104}, 019 (2011)
  [arXiv:1007.2654 [hep-th]].
  
\bibitem{GBounce}
T.~Qiu, J.~Evslin, Y.~F.~Cai, M.~Li and X.~Zhang,
  ``Bouncing Galileon Cosmologies,''
  JCAP {\bf 1110}, 036 (2011)
  [arXiv:1108.0593 [hep-th]];\\
D.~A.~Easson, I.~Sawicki and A.~Vikman,
  ``G-Bounce,''
  JCAP {\bf 1111}, 021 (2011)
  [arXiv:1109.1047 [hep-th]].

\bibitem{BMS}
R.~H.~Brandenberger, V.~F.~Mukhanov and A.~Sornborger,
  ``A Cosmological theory without singularities,''
  Phys.\ Rev.\  D {\bf 48}, 1629 (1993)
  [arXiv:gr-qc/9303001].

\bibitem{Biswas}
T.~Biswas, A.~Mazumdar and W.~Siegel,
  ``Bouncing universes in string-inspired gravity,''
  JCAP {\bf 0603}, 009 (2006)
  [arXiv:hep-th/0508194].

\bibitem{HLbounce}
R.~Brandenberger,
  ``Matter Bounce in Horava-Lifshitz Cosmology,''
  Phys.\ Rev.\  D {\bf 80}, 043516 (2009)
  [arXiv:0904.2835 [hep-th]].
  
\bibitem{Horava}
 P.~Horava,
  ``Quantum Gravity at a Lifshitz Point,''
  Phys.\ Rev.\  D {\bf 79}, 084008 (2009)
  [arXiv:0901.3775 [hep-th]].
  
\bibitem{LQC}
M.~Bojowald,
  ``Quantum Cosmology,''
  Lect.\ Notes Phys.\  {\bf 835}, 1 (2011);\\
  A.~Ashtekar and P.~Singh,
  ``Loop Quantum Cosmology: A Status Report,''
  Class.\ Quant.\ Grav.\  {\bf 28}, 213001 (2011)
  [arXiv:1108.0893 [gr-qc]].

\bibitem{emergent}
G.~F.~R.~Ellis and R.~Maartens,
  ``The emergent universe: Inflationary cosmology with no singularity,''
  Class.\ Quant.\ Grav.\  {\bf 21}, 223 (2004)
  [gr-qc/0211082];\\
G.~F.~R.~Ellis, J.~Murugan and C.~G.~Tsagas,
  ``The Emergent universe: An Explicit construction,''
  Class.\ Quant.\ Grav.\  {\bf 21}, 233 (2004)
  [gr-qc/0307112].
  
\bibitem{Tirtho2}
T.~Biswas, R.~Brandenberger, A.~Mazumdar and W.~Siegel,
  ``Non-perturbative gravity, Hagedorn bounce and CMB,''
  JCAP {\bf 0712}, 011 (2007)
  [arXiv:hep-th/0610274].

\bibitem{NBV}
A.~Nayeri, R.~H.~Brandenberger and C.~Vafa,
  ``Producing a scale-invariant spectrum of perturbations in a Hagedorn  phase
  of string cosmology,''
  Phys.\ Rev.\ Lett.\  {\bf 97}, 021302 (2006)
  [arXiv:hep-th/0511140].

\bibitem{Perlt}
 J.~Kripfganz and H.~Perlt,
  ``Cosmological Impact Of Winding Strings,''
  Class.\ Quant.\ Grav.\  {\bf 5}, 453 (1988).

\bibitem{RHBSGrev}
R.~H.~Brandenberger,
  ``String Gas Cosmology,''
  arXiv:0808.0746 [hep-th].

\bibitem{BWrev}
T.~Battefeld and S.~Watson,
  ``String gas cosmology,''
  Rev.\ Mod.\ Phys.\  {\bf 78}, 435 (2006)
  [arXiv:hep-th/0510022].

\bibitem{Hagedorn}
R.~Hagedorn,
  ``Statistical Thermodynamics Of Strong Interactions At High-Energies,''
  Nuovo Cim.\ Suppl.\  {\bf 3}, 147 (1965).

\bibitem{Pol}
J. Polchinski, \textit{String Theory, Vols. 1 and 2},
(Cambridge Univ. Press, Cambridge, 1998).

\bibitem{Boehm}
T.~Boehm and R.~Brandenberger,
  ``On T-duality in brane gas cosmology,''
  JCAP {\bf 0306}, 008 (2003)
  [arXiv:hep-th/0208188].

\bibitem{Mairi}
 M.~Sakellariadou,
  ``Numerical Experiments in String Cosmology,''
  Nucl.\ Phys.\  B {\bf 468}, 319 (1996)
  [arXiv:hep-th/9511075].

\bibitem{Cleaver}
G.~B.~Cleaver and P.~J.~Rosenthal,
  ``String cosmology and the dimension of space-time,''
  Nucl.\ Phys.\  B {\bf 457}, 621 (1995)
  [arXiv:hep-th/9402088].

\bibitem{Col2}
R.~Easther, B.~R.~Greene, M.~G.~Jackson and D.~Kabat,
  ``String windings in the early universe,''
  JCAP {\bf 0502}, 009 (2005)
  [arXiv:hep-th/0409121].

\bibitem{Danos}
R.~Danos, A.~R.~Frey and A.~Mazumdar,
  ``Interaction rates in string gas cosmology,''
  Phys.\ Rev.\ D {\bf 70}, 106010 (2004)
  [arXiv:hep-th/0409162].

\bibitem{Kabat}
B.~Greene, D.~Kabat and S.~Marnerides,
  ``Dynamical Decompactification and Three Large Dimensions,''
  arXiv:0908.0955 [hep-th].

\bibitem{Watson}
S.~Watson and R.~Brandenberger,
  ``Stabilization of extra dimensions at tree level,''
  JCAP {\bf 0311}, 008 (2003)
  [arXiv:hep-th/0307044];\\
S.~Watson,
  ``Moduli stabilization with the string Higgs effect,''
  Phys.\ Rev.\ D {\bf 70}, 066005 (2004)
  [arXiv:hep-th/0404177].

\bibitem{Subodh1}
S.~P.~Patil and R.~Brandenberger,
  ``Radion stabilization by stringy effects in general relativity and  dilaton
  gravity,''
  Phys.\ Rev.\ D {\bf 71}, 103522 (2005)
  [arXiv:hep-th/0401037].

\bibitem{Subodh2}
S.~P.~Patil and R.~H.~Brandenberger,
  ``The cosmology of massless string modes,''
  JCAP {\bf 0601}, 005 (2006)
  [arXiv:hep-th/0502069].

\bibitem{Edna}
R.~Brandenberger, Y.~K.~Cheung and S.~Watson,
  ``Moduli stabilization with string gases and fluxes,''
  JHEP {\bf 0605}, 025 (2006)
  [arXiv:hep-th/0501032].

\bibitem{Frey}
R.~J.~Danos, A.~R.~Frey and R.~H.~Brandenberger,
  ``Stabilizing moduli with thermal matter and nonperturbative effects,''
  Phys.\ Rev.\  D {\bf 77}, 126009 (2008)
  [arXiv:0802.1557 [hep-th]];\\
S.~Mishra, W.~Xue, R.~Brandenberger and U.~Yajnik,
  ``Supersymmetry Breaking and Dilaton Stabilization in String Gas Cosmology,''
  arXiv:1103.1389 [hep-th].

\bibitem{RHBrev}
R.~H.~Brandenberger,
  ``Lectures on the theory of cosmological perturbations,''
  Lect.\ Notes Phys.\  {\bf 646}, 127 (2004)
  [arXiv:hep-th/0306071].

\bibitem{MFB}
V.~F.~Mukhanov, H.~A.~Feldman and R.~H.~Brandenberger,
  ``Theory of cosmological perturbations. Part 1. Classical perturbations. Part
  2. Quantum theory of perturbations. Part 3. Extensions,''
  Phys.\ Rept.\  {\bf 215}, 203 (1992).

\bibitem{Berera}
A.~Berera,
  ``Warm inflation,''
  Phys.\ Rev.\ Lett.\  {\bf 75}, 3218 (1995)
  [astro-ph/9509049].

 \bibitem{Starob0} 
A.~A.~Starobinsky,
  ``Spectrum of relict gravitational radiation and the early state of the
  universe,''
  JETP Lett.\  {\bf 30}, 682 (1979)
  [Pisma Zh.\ Eksp.\ Teor.\ Fiz.\  {\bf 30}, 719 (1979)].

\bibitem{Mukh2}
V.~F.~Mukhanov,
  ``Quantum Theory of Gauge Invariant Cosmological Perturbations,''
  Sov.\ Phys.\ JETP {\bf 67}, 1297 (1988)
  [Zh.\ Eksp.\ Teor.\ Fiz.\  {\bf 94N7}, 1 (1988)].
  
\bibitem{Mukh3}
V.~F.~Mukhanov,
  ``Gravitational Instability Of The Universe Filled With A Scalar Field,''
  JETP Lett.\  {\bf 41}, 493 (1985)
  [Pisma Zh.\ Eksp.\ Teor.\ Fiz.\  {\bf 41}, 402 (1985)].
  
\bibitem{Sasaki}
M.~Sasaki,
  ``Large Scale Quantum Fluctuations in the Inflationary Universe,''
  Prog.\ Theor.\ Phys.\  {\bf 76}, 1036 (1986).

\bibitem{Lyth0}
D.~H.~Lyth,
  ``Large Scale Energy Density Perturbations And Inflation,''
  Phys.\ Rev.\  D {\bf 31}, 1792 (1985).

\bibitem{BST} 
J.~M.~Bardeen, P.~J.~Steinhardt and M.~S.~Turner,
  ``Spontaneous Creation Of Almost Scale - Free Density Perturbations In An
  Inflationary Universe,''
  Phys.\ Rev.\  D {\bf 28}, 679 (1983);\\
R.~H.~Brandenberger and R.~Kahn,
  ``Cosmological Perturbations In Inflationary Universe Models,''
  Phys.\ Rev.\  D {\bf 29}, 2172 (1984).
  
\bibitem{BD}
N.~D.~Birrell and P.~C.~W.~Davies,
  ``Quantum Fields In Curved Space,''
{\it  Cambridge, Uk: Univ. Pr. ( 1982) 340p}

\bibitem{Sudarsky}
D.~Sudarsky,
  ``Shortcomings in the Understanding of Why Cosmological Perturbations Look Classical,''
  Int.\ J.\ Mod.\ Phys.\ D {\bf 20}, 509 (2011)
  [arXiv:0906.0315 [gr-qc]].

\bibitem{HZ2}
H. Zinkernagel,
``Some Trends in the Philosophy of Physics",
Theoria {\bf 71}, 215 (2011);\\
S. Rugh and H. Zinkernagel, 
``Weyl's Principle: Cosmic Time and Quantum Fundamentalism",
in D. Dirks et al. (eds), {\it Explanation, Prediction and Confirmation, The Philosophy
of Science in a European Perspective 2}, DOI 10.1007/978-94-007-1180-8\_28
(Springer, 2011).

\bibitem{Starob3}
C.~Kiefer, I.~Lohmar, D.~Polarski and A.~A.~Starobinsky,
  ``Pointer states for primordial fluctuations in inflationary cosmology,''
  Class.\ Quant.\ Grav.\  {\bf 24}, 1699 (2007)
  [arXiv:astro-ph/0610700].
  
\bibitem{Martineau}
 P.~Martineau,
  ``On the decoherence of primordial fluctuations during inflation,''
  Class.\ Quant.\ Grav.\  {\bf 24}, 5817 (2007)
  [arXiv:astro-ph/0601134].

\bibitem{ABB}
S.~Alexander, T.~Biswas and R.~H.~Brandenberger,
  ``On the Transfer of Adiabatic Fluctuations through a Nonsingular
  Cosmological Bounce,''
  arXiv:0707.4679 [hep-th].

\bibitem{Saremi}
R.~Brandenberger, H.~Firouzjahi and O.~Saremi,
  ``Cosmological Perturbations on a Bouncing Brane,''
  JCAP {\bf 0711}, 028 (2007)
  [arXiv:0707.4181 [hep-th]].

\bibitem{HLbounce2}
X.~Gao, Y.~Wang, W.~Xue and R.~Brandenberger,
  ``Fluctuations in a Ho\v{r}ava-Lifshitz Bouncing Cosmology,''
JCAP {\bf 1002}, 020 (2010)
  [arXiv:0911.3196 [hep-th]];\\
  X.~Gao, Y.~Wang, R.~Brandenberger and A.~Riotto,
  ``Cosmological Perturbations in Ho\v{r}ava-Lifshitz Gravity,''
 Phys.\ Rev.\ D {\bf 81}, 083508 (2010)
  [arXiv:0905.3821 [hep-th]].

\bibitem{bispectrum}
Y.~-F.~Cai, W.~Xue, R.~Brandenberger and X.~Zhang,
  ``Non-Gaussianity in a Matter Bounce,''
  JCAP {\bf 0905}, 011 (2009)
  [arXiv:0903.0631 [astro-ph.CO]].
  
\bibitem{processing}
R.~H.~Brandenberger,
  ``Processing of Cosmological Perturbations in a Cyclic Cosmology,''
  Phys.\ Rev.\  D {\bf 80}, 023535 (2009)
  [arXiv:0905.1514 [hep-th]].

\bibitem{BNPV2}
R.~H.~Brandenberger, A.~Nayeri, S.~P.~Patil and C.~Vafa,
  ``String gas cosmology and structure formation,''
  Int.\ J.\ Mod.\ Phys.\  A {\bf 22}, 3621 (2007)
  [arXiv:hep-th/0608121].

\bibitem{Betal}
R.~H.~Brandenberger {\it et al.},
  ``More on the spectrum of perturbations in string gas cosmology,''
  JCAP {\bf 0611}, 009 (2006)
  [arXiv:hep-th/0608186].

\bibitem{KKLM}
N.~Kaloper, L.~Kofman, A.~Linde and V.~Mukhanov,
  ``On the new string theory inspired mechanism of generation of  cosmological
  perturbations,''
  JCAP {\bf 0610}, 006 (2006)
  [arXiv:hep-th/0608200].

\bibitem{Deo}
N.~Deo, S.~Jain, O.~Narayan and C.~I.~Tan,
  ``The Effect of topology on the thermodynamic limit for a string gas,''
  Phys.\ Rev.\  D {\bf 45}, 3641 (1992).

\bibitem{Ali}
A.~Nayeri,
   ``Inflation free, stringy generation of scale-invariant cosmological
  fluctuations in D = 3 + 1 dimensions,''
  arXiv:hep-th/0607073.

\bibitem{BNPV1}
R.~H.~Brandenberger, A.~Nayeri, S.~P.~Patil and C.~Vafa,
  ``Tensor modes from a primordial Hagedorn phase of string cosmology,''
  Phys.\ Rev.\ Lett.\  {\bf 98}, 231302 (2007)
  [arXiv:hep-th/0604126].

\end{thebibliography}
\end{document}